\documentclass[sn-mathphys]{sn-jnl}

\usepackage[utf8]{inputenc}
\usepackage{amsmath,amssymb}
\usepackage{outlines}
\usepackage{subcaption}
\usepackage{mathtools}
\usepackage{graphicx}
\usepackage{adjustbox}
\usepackage{booktabs}

\jyear{2022}%

\theoremstyle{thmstyleone}%
\theoremstyle{thmstyletwo}%

\theoremstyle{thmstylethree}%

\begin{document}

\title[Chaotic diffusion in the Second Fundamental Model of Resonance]
{\centering 
Semi-analytical estimates for the chaotic diffusion in the Second Fundamental Model of Resonance. Application to Earth's navigation satellites
}


\author*[1]{\fnm{Edoardo} \sur{Legnaro}}\email{legnaro@academyofathens.gr}

\author[2]{\fnm{Christos} \sur{Efthymiopoulos}}\email{cefthym@math.unipd.it}

\author[3]{\fnm{Maria} \sur{Harsoula}}\email{mharsoul@academyofathens.gr}

\affil*[1]{
    \orgdiv{Department of Physics}, \orgname{Aristotle University of Thessaloniki}, \orgaddress{\city{Thessaloniki}, \postcode{54124}, \country{Greece}}\\
    \orgdiv{Research Center for Astronomy and Applied Mathematics}, \orgname{Academy of Athens}, \orgaddress{\city{Athens}, \postcode{11527}, \country{Greece}} \\
    \orgdiv{Department of Mathematics}, \orgname{Università degli Studi di Padova}, \orgaddress{\street{Via Trieste 63}, \city{Padova}, \postcode{35121}, \country{Italy}}
    }

\affil[2]{\orgdiv{Department of Mathematics}, \orgname{Università degli Studi di Padova}, \orgaddress{\street{Via Trieste 63}, \city{Padova}, \postcode{35121}, \country{Italy}}}

\affil[3]{\orgdiv{Research Center for Astronomy and Applied Mathematics}, \orgname{Academy of Athens}, \orgaddress{\city{Athens}, \postcode{11527}, \country{Greece}}}


\abstract{We discuss the applicability of the Melnikov and Landau-Teller theories in obtaining semi-analytical estimates of the speed of chaotic diffusion in systems driven by the separatrix-like stochastic layers of a resonance belonging to the `second fundamental model' (SFM)\cite{henrard1983second}. Stemming from the analytic solution for the SFM in terms of Weierstrass elliptic functions, we introduce stochastic Melnikov and Landau-Teller models allowing to locally approximate chaotic diffusion as a sequence of uncorrelated `jumps' observed in the time series yielding the slow evolution of an ensemble of trajectories in the space of the adiabatic actions of the system. Such jumps occur in steps of one per homoclinic loop. We show how a semi-analytical determination of the probability distribution of the size of the jumps can be arrived at by the Melnikov and Landau-Teller approximate theories. Computing also the mean time required per homoclinic loop, we arrive at estimates of the chaotic diffusion coefficient in such systems. As a concrete example, we refer to the long-term diffusion of a small object (e.g. Earth navigation satellite or space debris) within the chaotic layers of the so-called $2g+h$ lunisolar resonance, which is of the SFM type. After a suitable normal form reduction of the Hamiltonian, we compute estimates of the speed of diffusion of these objects, which compare well with the results of numerical experiments. }




\maketitle

\section{Introduction}
In the present paper, we deal with the characterization of the chaotic diffusion in N-degree of freedom nonlinear Hamiltonian dynamical systems of the form:
\begin{equation}\label{eq:hamgen}
H(X,Y,\varphi,J)=H_{res}(X,Y)+H_a(J,\varphi) + \varepsilon H_c(X,Y,\varphi,J)~~.  
\end{equation}
System (\ref{eq:hamgen}) involves a resonant degree of freedom (variables $(X,Y)$) coupled to oscillators described by the $N-1$ pairs of action-angle variables $J\in\mathbb{R}^{N-1}$, $\varphi\in\mathbb{T}^{N-1}$. We require that: 

i) $H_{res}(X,Y)$ be a one-degree of freedom Hamiltonian belonging to the class called (in \cite{henrard1983second}) the `second fundamental model of resonance'. This implies the existence of a separatrix whose form is different from the one of the pendulum, emanating from the origin $(X,Y)=(0,0)$, which is assumed to be a hyperbolic equilibrium point. Here we focus on the case of a `figure-8' separatrix, characteristic of nonlinear systems with a quartic correction to a harmonic repulsor model $X^2-Y^2$. However, the results presented below are straightforward to generalize in practically any resonant model $H_{res}$ of the SFM type. 

ii) the Hamiltonian $H_a$ be of the form $H_a=\omega\cdot J+H_{1,a}(J,\varphi)$, with $H_{1,a}$ small and periodic in the angles $\varphi$. Thus, under the flow of $H_a$ the `actions' $J$ are quasi-conserved, and the angles $\varphi$ evolve quasi-linearly, approximately with the $N-1$ frequencies $\omega\in\mathbb{R}^{N-1}$, which are assumed to satisfy no commensurability condition. 

iii) We assume $\varepsilon$ small, $\mid\varepsilon\mid<<1$, while $H_c$ is assumed to have a polynomial dependence on $(X,Y)$ of degree second or higher. This implies that, for $\varepsilon\neq 0$, any trajectory with initial condition $(X(t=0),Y(t=0))=(0,0)$ remains always confined to the $(2N-2)-$dimensional sub-manifold:
\begin{equation}\label{cmanif}
\mathcal{C}_0\subset\mathbb{R}^{N-1}\times\mathbb{T}^{N-1}:
\left\{(X,Y)=(0,0),~J\in\mathcal{D}\subseteq\mathbb{R}^{N-1},~
\varphi\in\mathbb{T}^{N-1}\right\}~~.
\end{equation}   
Hence, $\mathcal{C}_0$, called hereafter the `center manifold', is invariant under the flow of full Hamiltonian (\ref{eq:hamgen}).    

Since $H_a$ is nearly integrable, averaging theory can be used to obtain $N-1$ quasi-integrals $\tilde{J}$ along any trajectory. 
When the coupling term $\varepsilon H_c$ is taken into account, however, we find a slow ($\mathcal{O}(\varepsilon)$) variation of the quasi-integrals $\tilde{J}$, namely $\dot{\tilde{J}}=\mathcal{O}(\varepsilon)$. 
The modelization of the long-term evolution of $\tilde{J}$ under the effects of the coupling term $H_c$ is a classical problem, related also to the so-called problem of the `Arnold diffusion' (see \cite{eftpaez2023} for an extended list of references).
Roughly speaking, it is known that the evolution of the adiabatic actions $\tilde{J}$ is not uniform in time, but proceeds essentially by a sequence of discrete `jumps'. 
Each individual jump corresponds to a `homoclinic pulse', i.e., a quick change in the values $X(t),Y(t)$ produced recurrently. 
As these variables repeatedly depart from the neighborhood of the manifold ${\cal C}_0$, they move in orbits which essentially `shadow' the homoclinic pulses of the separatrices of the resonant model $H_r$ and then return again to the neighborhood of ${\cal C}_0$. 
This process has been detailed in \cite{chirikov1979universal}, in the case when $H_{res}$ is given by the pendulum model. In particular, predictions on the size of the jump can be obtained by the so-called `whisker' map (see \cite{chirikov1979universal}), obtained in turn from the coupling function $H_c$. Its role in the process of Arnold diffusion is set on rigorous ground in \cite{chierchia1994drift}. 
A description of the recurrent excursions of the trajectories far from, and back to, the neighborhood of ${\cal C}_0$ can be done in the framework of the `scattering map' theory (see \cite{delshams2006geometric}, \cite{delshams2008geometric}). 
Furthermore, for $\varepsilon$ small, the jumps can be estimated on the basis of the remainder terms of a normal form construction aiming to eliminate the angles $\varphi$ up to an optimal order, in which the remainder becomes the least possible. This process was recently examined in \cite{guzzo2020semi}, in relation to the problem of Arnold diffusion (\cite{arnold1964instability}; see references in \cite{guzzo2020semi} and also in the review \cite{eftpaez2023}). 

In most works on the subject, as those cited in the references above, the model chosen to represent the hyperbolic degree of freedom is the pendulum Hamiltonian, which, according to the terminology introduced in \cite{henrard1983second}, is the archetype model for describing resonances of the so-called `first fundamental model'. In most problems of celestial mechanics or plasma physics, instead, we encounter resonances of the `second fundamental model', which the present paper focuses on. 

A main aspect of our present study is to investigate the connections between stochastic (random phase) models of diffusion based on one hand on the \textit{Melnikov} approach (see \cite{eftpaez2023}), and, on the other, on the implementation of the \textit{Landau-Teller} theory to such systems. By Landau-Teller theory we refer to a heuristic approach proposed in \cite{LandauTeller1936} in the context of the theory of sound dispersion, based on the original idea found back in \cite{jeans1903xxxv} and \cite{jeans1905xi}. The theory was revised in \cite{rapp1960complete}, \cite{benettin1993landau} and \cite{benettin1997conservation}, being now applied to the study of the evolution of adiabatic invariants in nearly integrable Hamiltonian systems. It basically consists of modeling the homoclinic pulses corresponding to near-separatrix motions of the hyperbolic variables $(X,Y)$ via a Fourier representation, giving rise to a model $(\tilde{X}(t),\tilde{Y}(t))$ for the pulse. In the spirit of the Melnikov approximation, this model is then introduced in the equations of motion for the adiabatic actions, leading to a Fourier representation of the right-hand side of these equations. 
Semi-analytical estimates on the variations of the action variables can then be obtained on the basis of the above formalism. One main point emphasized below is the fact that when the Hamiltonian $H_{1,a}$ is negligible (as it happens in many applications), the Landau-Teller theory, in the form described above, essentially leads to an \textit{integrable} model (except for a possibly exponentially small remainder), which implies no long-term chaotic diffusion in timescales comparable to the main Lyapunov times of the system. We propose that this can be remedied by introducing a \textit{stochastic Landau-Teller} model of diffusion. On the other hand, instead of the machinery of deterministic whisker or separatrix maps of Melnikov's theory, one can also introduce stochastic models based on a random phase approximation, as pointed out already in \cite{chirikov1979universal}. We discuss below how these different modellings of the chaotic diffusion within the stochastic layers of the resonance compare and/or are related to each other. 
  
In the sections to follow we first examine an `archetype model', i.e. a toy example of a Hamiltonian of the form (\ref{eq:hamgen}), from which three different levels of modelling of the chaotic diffusion can be obtained: in the most basic level, we numerically compare the diffusion of the adiabatic actions as observed in the exact flow and in the `Melnikov approximation', i.e. after the substitution of the pulse model $(\tilde{X}(t),\tilde{Y}(t))$ in the equations of motion. In the second level, we obtain a semi-analytic estimate of the size of the jumps in the adiabatic actions based on an integrable (Landau-Teller) approximation to the Hamiltonian, valid locally, along just one homoclinic pulse. In the third level, we assume a stochastic process in which the coefficients of the locally integrable model can be varied at each loop according to random choice from a model probability distribution function obtained through estimates valid within the thin chaotic layer surrounding the separatrix of the theoretical model $H_{res}(X,Y)$.     

Finally, as an example of the applicability of the above methods we consider a dynamical system stemming from celestial mechanics, namely the one corresponding to the long-term diffusion in the space of orbital elements for a small object (e.g. space debris) in the orbital domain corresponding to navigation satellites under the influence of lunisolar resonances in the Medium Earth Orbit (MEO) region. This topic is of particular current interest within the science/engineering community in astrodynamics, as lunisolar resonances can be exploited to tackle the urgent problem of the growing space debris population.
In fact, such resonances have the effect of increasing an object's eccentricity, possibly up until the orbit's perigee becomes small enough that the friction of the Earth's atmosphere will make the object re-enter (see \cite{rosengren2015chaos}, \cite{daquin2016dynamical}
\cite{celletti2016study} \cite{gkolias2016order}, \cite{rossi2018redshift}, and
\cite{celletti2017analytical}).
In \cite{daquin2021deep} and \cite{Legnaro_Efthymiopoulos_2022}, an analytic theory is given for these secular lunisolar resonances that also provides insights into the eccentricity growth phenomenon. Here, instead, we discuss the application of our proposed semi-analytical methods for the quantification of the speed of diffusion along such resonances.

The structure of the paper is the following.
Section \ref{sec:theory} presents an archetype model for diffusion under the regime of the `second fundamental model of resonance'. This model allows to exemplify one by one all the intermediate steps and models of increasing complexity leading to our semi-analytical estimates on the rate of the chaotic diffusion.  
In Section \ref{sec: diffusion_2g+h} we apply the theory to semi-analytically predict the diffusion coefficient for space debris motions along the so-called $2g+h$ lunisolar resonance for Earth navigation satellites. Results will be presented for the particular cases of GPS and Galileo satellites. Section \ref{sec:conclusions} summarizes the basic conclusions from the present study. 

\section{An Archetype Model}
\label{sec:theory}

\subsection{Hamiltonian}
As an archetype model for our study, we consider the 3DOF Hamiltonian
 \begin{equation}
    \label{eq:HB}
    H = - S \cos 2 \sigma + S^2 \lambda\left[ 1 
    + \varepsilon\cos(2\sigma - \phi) 
    + \varepsilon\cos(2\sigma - \phi_2)\right] 
    + A \Omega+ A_2 \Omega_2
 \end{equation}
where $(A,A_2)\in\mathbb{R}^2$, $S \geq 0 $, $\sigma \in \mathbb{S}^1$, and $\varepsilon, \Omega, \Omega_2 \geq 0$. This system consists of two oscillators (action-angle variables $(A,\phi)$, $(A_2,\phi_2)$) coupled with a so-called `resonant' degree of freedom $(S,\sigma)$. The term `resonant' refers to the fact that in the uncoupled case $\varepsilon=0$ the integrable flow in the variables $(S,\sigma)$ produces a phase-portrait belonging to the class of the `second fundamental resonance' (SFM2 model; see \cite{henrard1983second}). 

Introducing Poincaré canonical variables through the canonical transformation $X= \sqrt{2 S} \sin \sigma$, 
$Y = \sqrt{2 S}\cos \sigma$ yields
\begin{eqnarray}
    \label{eq:HPoinc}
    H&=& \frac{1}{2} \left( X^2 - Y^2 \right) + \frac{\lambda}{4} \left( X^2 + Y^2 \right)^2 + A \Omega + A_2 \Omega_2 \\ ~~\nonumber
    &+& \frac{\varepsilon \lambda}{2}  \, X Y \left(X^2 + Y^2\right)  \left(\sin \phi + \sin \phi_2\right) -  \frac{\varepsilon \lambda}{4}    \left(X^4 - Y^4\right) \left(\cos \phi + \cos \phi_2\right)
\end{eqnarray}

From Hamilton's equations, for $X=Y=0$ we obtain $\dot{X}=\dot{Y}=0$ by identity. Thus, all initial conditions $\left( A(0),A_2(0),\phi(0),\phi_2(0) \right)$ with $X(0)=Y(0)=0$ lie on hyperbolic 2D-tori ${\cal T}(A,A_2)$, in which $A(t)=A(0)=const$, $A_2(t)=A_2(0)=const$, while the angles evolve linearly, $\phi(t)=\phi(0)+\Omega t$, $\phi_2(t)=\phi_2(0)+\Omega_2 t$. 

\subsection{Numerical probes of chaotic diffusion}
\label{sec:numprobes}
We are now interested in understanding the long-term evolution of ensembles of initial conditions chosen near any member of the set of hyperbolic tori ${\cal T}(A,A_2)$. We can observe that the orbits obtained for different initial values $A(0)$ and $A_2(0)$ are identical as regards the evolution of all remaining variables. Thus, without loss of generality, we can choose the torus ${\cal T}_0={\cal T}(A=0,A_2=0)$ as representative, and focus our study to the evolution of ensembles of initial conditions taken in the neighborhood of ${\cal T}_0$. 

\begin{figure}
\centering
\includegraphics[width=0.7\textwidth]{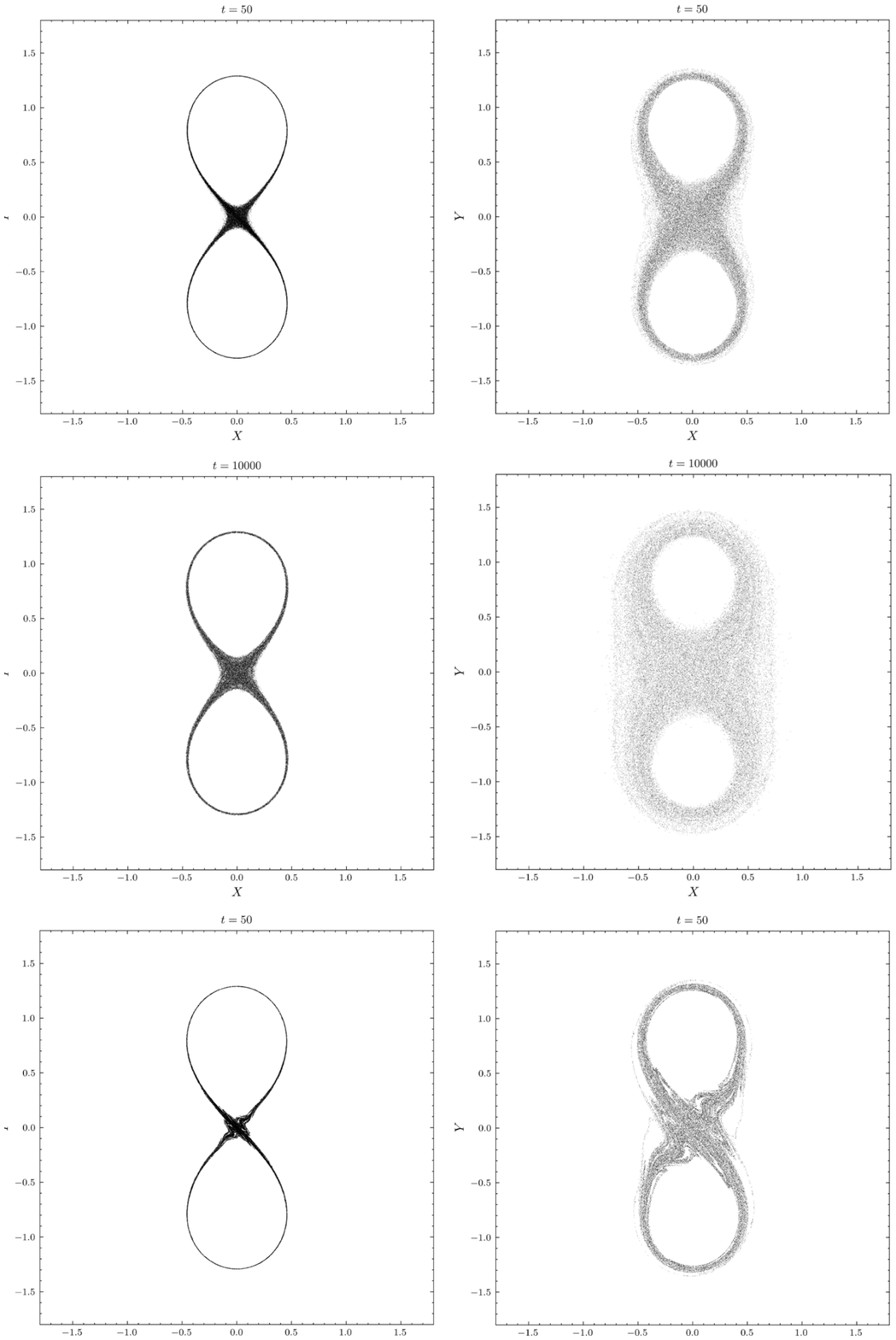}
\caption{\small The $(X,Y)$ projection of the iterates of 50000 initial conditions of the archetype model for $\lambda=1.2$, $\Omega=5$, $\Omega_2=5\sqrt{2}$ and $\varepsilon=10^{-3}$ (left), or $\varepsilon=10^{-2}$ (right). The initial conditions are taken in the small square around the torus ${\cal C}_0$ $\left[ -0.05, 0.05 \right]^2$. The iterates of all trajectories are collected after the times $t=50$ (top) and $t=10000$ (middle). Bottom: the iterates of the same initial conditions after $t=50$ in the case of the 2DOF Hamiltonian model of Eq.(\ref{eq:HPoinc2D}).}
    \label{fig:pmap3D}
    \end{figure}
Figure \ref{fig:pmap3D} shows the typical form of the phase portrait for the Hamiltonian (\ref{eq:HPoinc}). The projection in the $(X,Y)$ plane of the orbit's evolution is shown for the parameters $\lambda=1.2$, $\Omega=5$, $\Omega_2=5\sqrt{2}$,  and $\varepsilon = 10^{-3}$ (left column) or $\varepsilon=10^{-2}$ (right column). In each case, we consider initial conditions taken (besides $A(0)=A_2(0)=0$) by setting $\phi=\phi_2=0$, while selecting $50000$ initial conditions in a square box $X=(-L,L)$, $Y=(-L,L)$ around the hyperbolic torus (corresponding to $X=Y=0$), with $L=0.05$. For both values of $\varepsilon$ we distinguish the formation of a chaotic layer, with the points distributed around a figure-8 separatrix of the `resonant' integrable model $H_r$ given by 
\begin{equation}
    \label{eq:Hres}
    H_r=H_{\varepsilon=0} =  
    \frac{1}{2} \left( X^2 - Y^2 \right) + \frac{1}{4} \lambda \left( X^2 + Y^2 \right)^2 + A \Omega + A_2 \Omega_2~~.
\end{equation}    

\begin{figure}
\centering
\includegraphics[width=\textwidth]{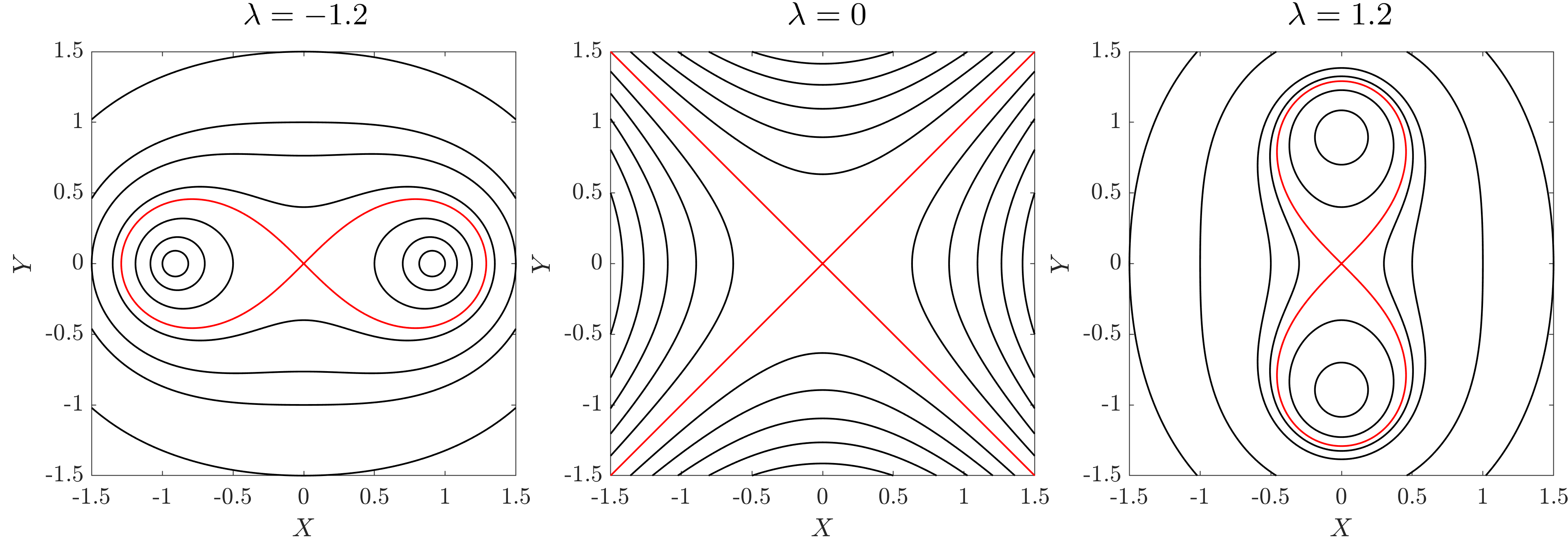}
\caption{\small Phase portrait of the integrable Hamiltonian $H_r$ when varying the parameter $\lambda$.}
\label{fig:bifurcations}
\end{figure}

Note that the phase space of the integrable model $H_r$ is the product of the plane 
${\cal P}_{X,Y}=(X,Y)$ with the 4D space ${\cal D}=\mathbb{R}^2\times\mathbb{T}^2$: 
the latter is foliated by invariant tori $\tau(A,A_2)$, labeled by any different pair of 
values $(A,A_2)$ which remain constant along any trajectory on the torus $\tau(A,
A_2)$. 
In particular, we have ${\cal T}(A,A_2)=(0,0)\times\tau(A,A_2)$. 
Also, under the integrable model $H_r$ the phase space structure in ${\cal P}_{X,Y}$ is as shown in Fig. \ref{fig:bifurcations}. The parameter $\lambda$ controls the position of the stable points (which are on the X-axis, for $\lambda<0$, or in the Y-axis, for $\lambda>0$). In fact, the choice of this control parameter is motivated by the necessity to consider a simple model with phase-space structure similar to the one of the Hamiltonian models of the Kozai-Lidov type applying to the case of lunisolar resonances for satellite orbits, as discussed in the next section. Similarly, the choice of values for the parameter $\varepsilon$ is motivated by comparison with the width of the stochastic layer in the case of navigation satellite orbits (GPS and Galileo, respectively, see next section).
However, we emphasize that similar phase portraits are obtained in most problems of secular resonances encountered in Celestial Mechanics.

Figure \ref{fig:pmap3D} offers a first insight into the type of chaotic diffusion exhibited by trajectories of the system (\ref{eq:HPoinc}) with initial conditions close to the hyperbolic torus ${\cal T}_0$. The diffusion manifests itself as a slow spreading of the orbits in a figure-8 shaped chaotic layer whose width grows in time. A noteworthy remark regards the existence of some partial barriers to chaotic diffusion, generated by the stickiness around possible invariant tori of the full 3DOF Hamiltonian. 
The existence of such tori, which act as partial barriers to diffusion (see, for example, \cite{efthymiopoulos1997stickiness}, \cite{contopoulos1999destruction} and \cite{contopoulos2008stickiness}) can be inferred from examining a model similar to (\ref{eq:HPoinc}), but with only two DOF, namely 
\begin{eqnarray}
    \label{eq:HPoinc2D}
    H_{2DOF} &=& \frac{1}{2} \left( X^2 - Y^2 \right) + \frac{\lambda}{4} \left( X^2 + Y^2 \right)^2 + A \Omega\\
    &-& {\varepsilon\lambda\over 4} \left(X^4-Y^4\right)\cos\phi 
    + {\varepsilon\lambda\over 2}\left(X^2+Y^2\right)XY\sin\phi 
    ~~\nonumber .
\end{eqnarray}
The last row of Figure \ref{fig:pmap3D} shows the distribution of the Poincar\'{e} map subsequents for the same initial conditions as in Fig.\ref{fig:pmap3D}, but in the case of the system (\ref{eq:HPoinc2D}). Note that, now, the existence of Kolmogorov-Arnold-Moser (KAM) tori (both librational, around each of the two symmetric stable periodic orbits surrounded by the figure-8 chaotic layer, or rotational, i.e., surrounding the layer) provides a total barrier that cannot be crossed by chaotic orbits. 

In the 3D case, instead, any 3D-tori of the KAM type cannot separate topologically the regions occupied by chaotic orbits within any 5-dimensional manifold of constant energy. As the top and middle panels of Fig.\ref{fig:pmap3D} show, however, the chaotic orbits in the 3DOF case also exhibit some difficulty in passing through the various partial barriers possibly created by the existence of 3D invariant tori in this region.

Finally, in the 2DOF case, we notice the presence of secondary islands of stability embedded in the stochastic layer, where the chaotic orbits cannot penetrate. Such islands are known to generate long-term correlations of the `phases' (the angular variables of the problem) due to `stickiness' effects (see \cite{contopoulos2010stickiness} for a review).  

\begin{figure}
\centering
\includegraphics[width=1.0\textwidth]{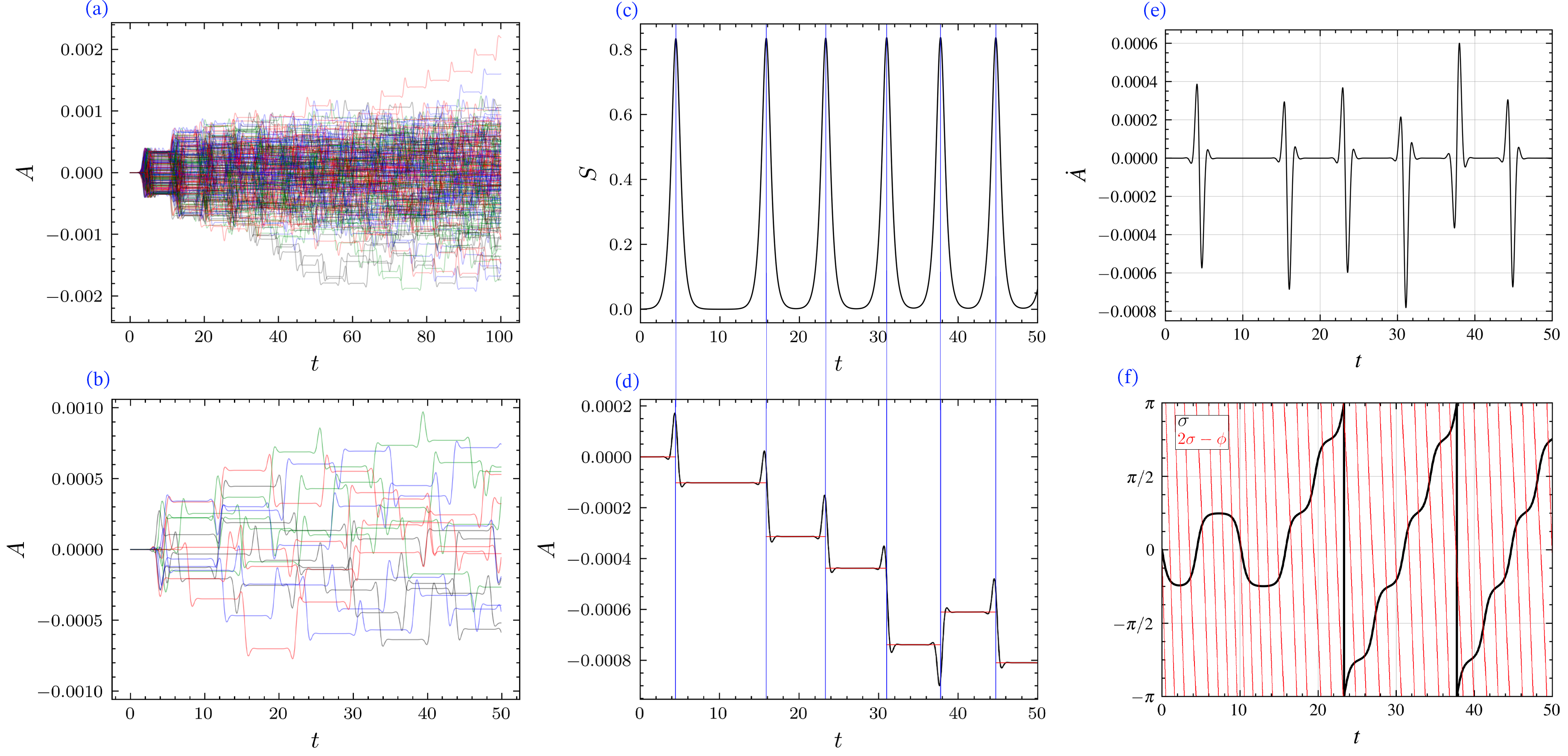}
\caption{(a) Evolution of $A(t)$ for a swarm of 250 orbits started all with $A(0)=A_2(0)=0$ and the remaining initial conditions closed to the hyperbolic torus ${\cal T}_0$ (see text), for $\varepsilon=0.001$, up to the time $t=100$. (b) A focus on few trajectories up to $t=50$. (c) Evolution of $S(t)$ for one trajectory along six consecutive homoclinic loops. (d) Detail of the evolution of $A(t)$ for the same trajectory as in (c). We observe that the jumps in the value of $A(t)$ occur in the middle of each homoclinic loop (when $S(t)$ reaches a local maximum). (e) $\dot{A}(t)$ for the same trajectory and in the same time interval. (f) evolution up to $t=50$ of the angle $\sigma(t)$ (black) and $2 \sigma(t) - \phi(t)$ (red).}
\label{fig:swarm}
\end{figure}

From now on we focus on the question of how the chaotic spreading of the orbits within the figure-8 stochastic layer described above influences the evolution of the `adiabatic' action variables $A(t)$ and $A_2(t)$. Some numerical insight into this question is provided by Fig.\ref{fig:swarm}. Panel (a) shows the evolution of $A(t)$ for a limited sub-sample of 250 orbits from the set of the orbits of Fig.(\ref{fig:pmap3D}), hence, all starting with $A(0)=0$, for $\varepsilon=0.001$. The evolution is shown up to $t=100$, while panel (b) shows a focus for fewer trajectories up to $t=50$. Note that similar remarks as those holding for the evolution of $A(t)$ can be made as regards the second adiabatic action $A_2(t)$. 

\begin{figure}
    \centering
    \includegraphics[width=0.8\textwidth]{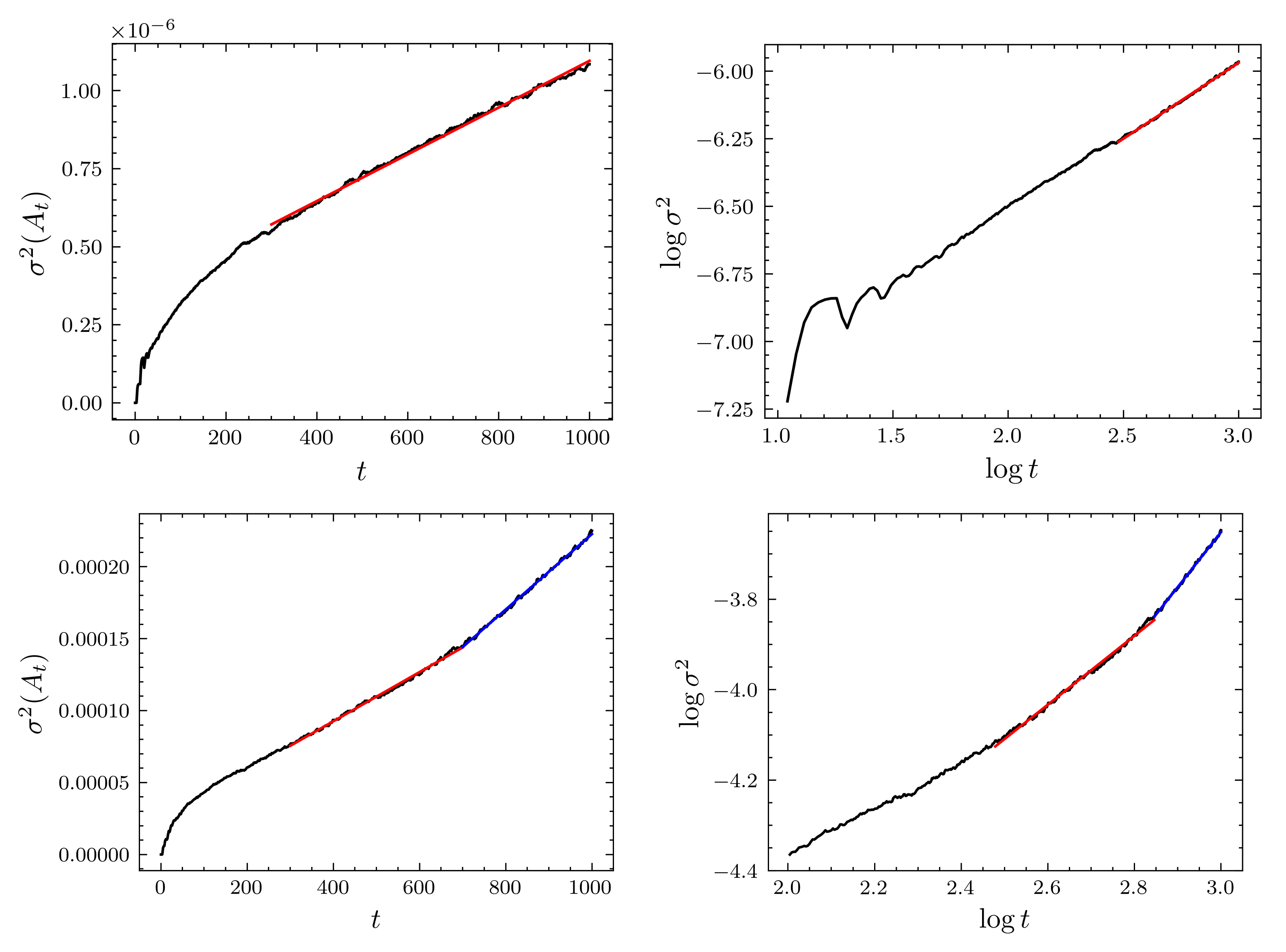}
    \caption{\small The growth of the dispersion $\sigma^2$ of the values of $A$ for the ensemble of 10000 simulated trajectories as a function of time shown in linear scale (left) or logarithmic scale (right) for $\varepsilon=0.001$ (top) and $\varepsilon=0.01$ (bottom). The linear fit in the data (red) yields a diffusion coefficient $D=7.5\times 10^{-10}$ in the interval $300\leq t\leq 1000$ for $\varepsilon=0.001$, and $D=1.7\times 10^{-7}$ in the interval $300\leq t\leq 700$ for $\varepsilon=0.01$.
    The log-log plots, instead, yield an effective sub-diffusive behavior $\sigma^2\propto t^a$ with $a=0.56$ and $a=0.76$ respectively. }
   \label{fig:diffusion}
\end{figure}
The main effect observed in Figs.\ref{fig:swarm}a,b is a systematic spreading of the values of $A$ for different trajectories as time goes on, which takes place through a sequence of abrupt `jumps' in the value of A of each trajectory, followed by intervals of time in which $A(t)$ temporarily stabilizes, forming a `plateau' of nearly constant value. 
A careful inspection of any single one of these orbits, as in panels (c) and (d) of Fig.\ref{fig:swarm}, shows that the jumps occur in a sequence of one per homoclinic loop in the plane $(X,Y)$. 
Namely, plotting the time-series $A(t)$ (Fig.\ref{fig:swarm}d) in comparison with the time-series of the resonant action variable $S(t)=\left( X(t)^2 + Y(t)^2 \right)/2$ (Fig.\ref{fig:swarm}c), one notes that the values of $A(t)$ remain practically constant as long as $S(t)$ remains close to zero, i.e., as long as the orbit remains close to the hyperbolic torus ${\cal T}_0$ which corresponds to the origin $(0,0)$ in the plane $(X,Y)$, while the jumps take place at the instant when the trajectory crosses the point along each homoclinic loop of largest distance ($S$ locally maximum) from the torus ${\cal T}_0$. 
Compared with Fig.\ref{fig:pmap3D}, the jumps in the adiabatic actions occur when the $(X,Y)$ projection of the trajectory is close to its locally uppermost, or lowermost point of the figure-8 shaped separatrix chaotic layer. Note also that the time series $S(t)$ itself shows a rather regular and recurrent behavior, i.e., starting from values close to zero it recurrently grows up to a maximum value, indicating that the corresponding orbit undergoes a large excursion away from ${\cal T}_0$, and then the orbit returns again to the neighborhood of ${\cal T}_0$, thus leading $S(t)$ again to values close to zero (Fig.\ref{fig:swarm}c). 
In fact, a crucial remark for what follows is that these loops yield an essentially identical form of the homoclinic pulse $S(t)$ around each maximum, the only irregularity concerning only the exact \textit{time separation} between any two consecutive pulses. 

Figure \ref{fig:diffusion} shows the growth in time of the spreading in the values of $A(t)$ up to $t=1000$ using a large ensemble of $N=10000$ initial conditions taken as in the previous figures. The dispersion is defined as 
\begin{equation}\label{eq:disp}
    \sigma^2(t)={1\over N}\sum_{i=1}^N A_i^2(t)
    -\left({1\over N}\sum_{i=1}^N A_i(t)\right)^2
\end{equation} 
where $A_i(t)$ is the value of the adiabatic action $A$ for the $i-th$ trajectory in the ensemble at time $t$. Quasi-linear diffusion coefficients in bounded time intervals can be estimated numerically by linear fitting as in the left panels of Fig.\ref{fig:diffusion}. On the other hand, plotting the same data in log-log scale suggests that the spreading of $A$ in time for ensembles of trajectories rather mimics a sub-diffusive process. Since, however, in the applications we are mostly interested in quasi-linear estimates on the speed of diffusion in the space of the adiabatic actions, in what follows we limit ourselves to a discussion of normal diffusion models in the framework of which estimates on local (linera) diffusion coefficients can be arrived at. In particular, we examine and quantify the degree to which the process of the spreading of the adiabatic actions via the `jumps' presented in Fig.\ref{fig:swarm} can be modelled as a stochastic process. To this end, we depart from the formalism based on the Melnikov or the Landau-Teller theories, to which we now turn our attention. 

\subsection{Estimates by Melnikov theory}
\label{sec:Melnikov}
\subsubsection{Melnikov Approximation}
Hamilton's equations for the adiabatic actions under the Hamiltonian (\ref{eq:HPoinc})) read:
\begin{equation}
    \label{eq:dotaa2}
    \dot{A}=-\frac{\partial H}{\partial \phi}= - \varepsilon \lambda S^{2} \sin (2\sigma-\phi),~~~
    \dot{A_2}=-\frac{\partial H}{\partial \phi_2}= - \varepsilon \lambda S^{2} \sin(2\sigma-\phi_2)~~.
\end{equation}
Melnikov's approximation consists in approximating $S(t)$, $\sigma(t)$, as well as $\phi(t)$, $\phi_2(t)$, with homoclinic pulse model $\hat{S}(t)$, $\hat{\sigma}(t)$, $\hat{\phi}(t)$, $\hat{\phi}_2(t)$ obtained by the separatrix solution under the integrable Hamiltonian $H_r(X,Y)$ for $t>t_0$.  
In the case of the adiabatic action $A(t)$, for example, this leads to:
\begin{equation}
    \label{eq:A_melnikov}
    A(t)\approx A(t_{0})-\int_{t_0}^{t} \varepsilon \lambda \hat{S}^{2}(u) \sin \left(2\hat{\sigma}(u)- \left( \phi(t_0)+\Omega u \right) \right) du.
\end{equation}
The approximation is based on the idea that the evolution of the resonant variables $(X,Y)$  (or $(S,\sigma)$) for initial conditions close to the separatrix of the integrable model is quite similar in the full model as in the integrable model. 
Figures \ref{fig:swarm}c,f show that this is essentially correct. On the other hand, we also require that in the r.h.s of the equations of motion the adiabatic actions be approximated as nearly constant,  while the angles canonically conjugated to the adiabatic actions ($(\phi,\phi_2)$ in our case) are approximated to have a linear evolution, e.g., $\hat{\phi}(t)=\phi(0)+\Omega t$. Note that this is in general the solution for the angles under the integrable part of the Hamiltonian $H_a$ of Eq.(\ref{eq:hamgen}), which, in our particular example, coincides with $H_a$, since $H_{1,a}=0$.  

\subsubsection{Solution for the Second Fundamental Model or Resonance}
The equations of motion of the resonant variables $(X,Y)$ under the integrable model $H_r$ are
\begin{align}\label{eq:eqmoxy}
\dot{X}&=\frac{\partial H}{\partial Y}=Y\left(-1+\lambda\left(X^{2}+Y^{2}\right)\right), \\
\dot{Y}&=-\frac{\partial H}{\partial X}=-X\left(1+\lambda\left(X^{2}+Y^{2}\right)\right)~~.
\end{align}
The equilibria are: $X=Y=0$ (unstable equilibrium at the origin) and 
$X=0$, $Y=\pm \sqrt{\frac{1}{\lambda}}$ for $\lambda>0$, or $Y=0$, $X=\pm \sqrt{-\frac{1}{\lambda}}$ for $\lambda<0$ (stable equilibria, corresponding to the `frozen orbits' in models of secular resonances, see next section). The phase portraits generated for $\lambda$ negative, zero or positive are shown in Fig.\ref{fig:bifurcations}.

For a particular value of the energy $E=S^2 \lambda - S \cos 2 \sigma$, the evolution of $S$ is given by $\dot{S}=-2 S \sin 2 \sigma$. Hence
\begin{equation}
    \sin 2 \sigma=-\sqrt{\frac{1}{S^{2}}\left(S^{2}-\left(E-\lambda S^{2}\right)\right)^{2}},
\end{equation}
so $\dot{S}=\sqrt{f(S)}$ with
$$f(S)=-4 E^{2}+4(1+2 E \lambda) S^{2}-4 \lambda^{2} S^{4}=a_{4}+4 a_{3} S+6
a_{2} S^{2}+4 a_{1} S^{3}+a_{0} S^{4}$$ with 
\begin{equation}
a_{4}=-4 E^{2}, \quad a_{3}=0,
\quad a_{2}=\frac{2}{3}(1+2 E \lambda), \quad a_{3}=0, \quad a_{4}=-4
\lambda^{2}.
\end{equation}

Following \cite{whittaker2020course}, we can find an analytic solution for $S$ as a function of the Weierstrass elliptic function $\oldwp$:
\begin{equation}
    \label{eq:g2}
    g_{2}=a_{0} a_{4}-4 a_{1} a_{3}+3 a_{2}^{2}=\frac{4}{3}\left(1+4 \lambda E+16
E^{2} \lambda^{2}\right),
\end{equation}
\begin{align}
    \begin{split}
    \label{eq:g3}
    g_{3} =& a_{0} a_{2} a_{4}+2 a_{1} a_{2} a_{3}-a_{2}^{3}-a_{0} a_{3}^{2}-a_{1}^{2} a_{4} \\
        =& \frac{8}{27}(1+2 E \lambda)(-1+4 E \lambda)(1+8 E \lambda)
    \end{split}
\end{align}

and 
$$
S(t)=S_{0}+\frac{1}{4} f^{\prime}\left(S_{0}\right) \cdot\left(\oldwp\left(t ;
g_{2}, g_{3}\right)-\frac{1}{24} f^{\prime
\prime}\left(S_{0}\right)\right)^{-1},
$$
$$
f'(S)=8 S\left(1+2 E \lambda-2 \lambda^{2} S^{2}\right), \quad f^{\prime
\prime}(S)=8\left(1+2 \lambda\left(E-3 \lambda s^{2}\right)\right) 
$$
\begin{equation}\label{eq:S}
S(t)=S_0+\dfrac{6 S_0 (1+2E \lambda -2S_0^2 \lambda^2 )}{-1-2E
\lambda +6 S_0^2 \lambda^2 +3 \oldwp(t;g_2,g_3)}.
\end{equation}
For the angle $\sigma$ we have $\cos 2 \sigma=\frac{\lambda S^{2}-E}{S}$, $\sin 2 \sigma=-\frac{1}{2} \frac{d}{d t}\log (S)$. Combining these equations we arrive at
\begin{equation}\label{eq:sigma}
2 \sigma(t)=\text{sgn}\left( -\frac{1}{2} \frac{d}{d t} \log (S(t)) \right) \arccos
\left( \frac{\lambda S(t)^{2}-E}{S(t)} \right).
\end{equation}
The above analytic solutions are periodic for all values $E\neq 0$, with a period deduced from the period of the Weirstrass elliptic function $ \oldwp $. Knowing $g_2$ and $g_3$, it is possible to find the two periods $\omega_1$ and $\omega_2$ of $\oldwp$ via the integrals:  
{\small \begin{align*}
\left\{ \omega_{1}\left(g_{2}, g_{3}\right), \omega_{3}\left(g_{2}, g_{3}\right)\right\}=\left\{\int_{e_{1}}^{\infty} \frac{1}{\sqrt{4 t^{3}-g_{2} t-g_{3}}} d t, i \int_{-\infty}^{e_{3}} \frac{1}{\sqrt{4 t^{3}-g_{2} t-g_{3}}} d t\right\},
\end{align*}
}%
with 
\begin{align}\label{eq:periods}
    & g_{2}, g_{3} \in \mathbb{R}, \; g_{2}^{3}-27 g_{3}^{2}>0, \\
    & 4 t^{3}-g_{2}t-g_{3}=4\left(t-e_{1}\right)\left(t-e_{2}\right)\left(t-e_{3}\right), \\
    & e_{1}>e_{2}>e_{3}
\end{align}
and finally $ \omega_2 = - \omega_1 - \omega_3 $. In our case, from
\eqref{eq:g2} and \eqref{eq:g3} it follows that $g_2$ and $g_3$ are functions of the
energy, while 
\begin{equation*}
    e_1= \frac{1}{3} (1-4 E \lambda), \; e_2= -\frac{2}{3} (1+2 E \lambda), \; e_3=\frac{1}{3} (1+8 E \lambda). 
\end{equation*}
The separatrix solution corresponds to the limit of Eqs.(\ref{eq:S}) and (\ref{eq:sigma}) when $E=0$. In the limit $E\rightarrow 0$ both periods deduced by the integrals (\ref{eq:periods}) tend to infinity. Without loss of generality, we treat the case $\lambda>0$. The initial condition $S_0=1/\lambda$ at $t=0$ leads to a pair of even symmetric solutions $S(t)=\pm S_\infty(t)$, $\sigma(t)=\sigma_\infty(t)(+\pi)$ with the property
\begin{equation}\label{eq:seplim} 
    \lim_{t\rightarrow -\infty}S_\infty(t)=
    \lim_{t\rightarrow \infty}S_\infty(t)=0,~
    \lim_{t\rightarrow -\infty}\sigma_\infty(t)=-{\pi\over 4},~
    \lim_{t\rightarrow \infty}\sigma_\infty(t)={\pi\over 4}~~.
\end{equation}

\subsubsection{Quantification of the size of jumps in the adiabatic actions}
Consider a trajectory with initial conditions at time $t=t_0$ very close to the torus ${\cal T}_0$. The time $t_0$ is in one of the intervals in which $S(t_0)$ is near a local minimum of the graph $S(t)$ (Fig.\ref{fig:swarm}c), while $A(t_0)$ is in an interval where the graph $A(t)$ (and analogously for $A_2(t)$) forms a local plateau. Let $t_{max}$ be the time in which $S(t)$ has its first local maximum after the time $t_0$. Define $\phi_0=\phi(t_0)+\Omega (t_{max}-t_0)$. Finally, let $t_1$ be the time of the next minimum of $S(t)$ after $t_0$. Following the formulas proposed in \cite{chirikov1979universal}, we then approximate the jump $\Delta A (\phi_0) =A(t_1)-A(t_0)$ between consecutive plateaus of $A(t)$ by the following Melnikov integral
\begin{equation}
    \Delta A \approx \Delta A_M = - \varepsilon \lambda \int_{-\infty}^{\infty}  \hat{S}_\infty^{2}(u) \sin \left(2\hat{\sigma}_\infty(u)- (\phi_{0}+\Omega u ) \right) du ~~.
    \label{eq:A_melnikov01}
\end{equation}
Note that, in generic models we have $\Delta A_M=\Delta A_M(\phi_0,\phi_{2,0})$, where the phase $\phi_{2,0}$ is defined analogously as $\phi_0$. 
Furthermore, since the half-period of any closed orbit $\hat{S}(t),\hat{\sigma}(t)$ tends to infinity close to the separatrix solution, we have in general $t_{max}-t_0>>2\pi/\Omega$ for trajectories sufficiently close to the solution $S_\infty(t),\sigma_\infty(t)$, implying that the quantity $\Omega (t_{max}-t_0)$ in the definition of the phase $\phi_0$ can end up (modulo $2\pi$) in practically any value in the interval $[0,2\pi)$, with uniform probability, independently of the value of $\phi(t_0)$. 
Let $\phi_0^{(j)}$, $\phi_{0,2}^{(j)}$, $j=1,2,\ldots$ be the sequence of phases computed as above for all the consecutive jumps as in Fig.\ref{fig:swarm}. Following \cite{chirikov1979universal}, the evolution of $A(t)$ can then be approximated as a discrete stochastic process defined by the sequence:
\begin{equation}\label{eq:darandomphi}
    A^{(j+1)}= A^{(j)}+\Delta A_M(\phi_{0}^{(j)},\phi_{0,2}^{(j)})
\end{equation}   
with the sequences $(\phi_{0}^{(j)},\phi_{0,2}^{(j)})$, each obtained by a random choice of values uniformly distributed in $[0,2\pi)$. While in the process (\ref{eq:darandomphi}) the phases of subsequent steps are assumed to be uncorrelated, the case of phase correlations is also dealt with in \cite{chirikov1979universal} by introducing a \textit{reduction factor} affecting all estimates on the speed of diffusion of the adiabatic actions given by the model. Such correlations can, for example, be generated due to stickiness effects within the separatrix chaotic layer, as discussed above (see \cite{cincotta2002arnold}, \cite{cincotta2014chirikov}; see also \cite{cincotta2023estimation} and references therein). 

\begin{figure}
    \centering
    \includegraphics[width=1.0\textwidth]{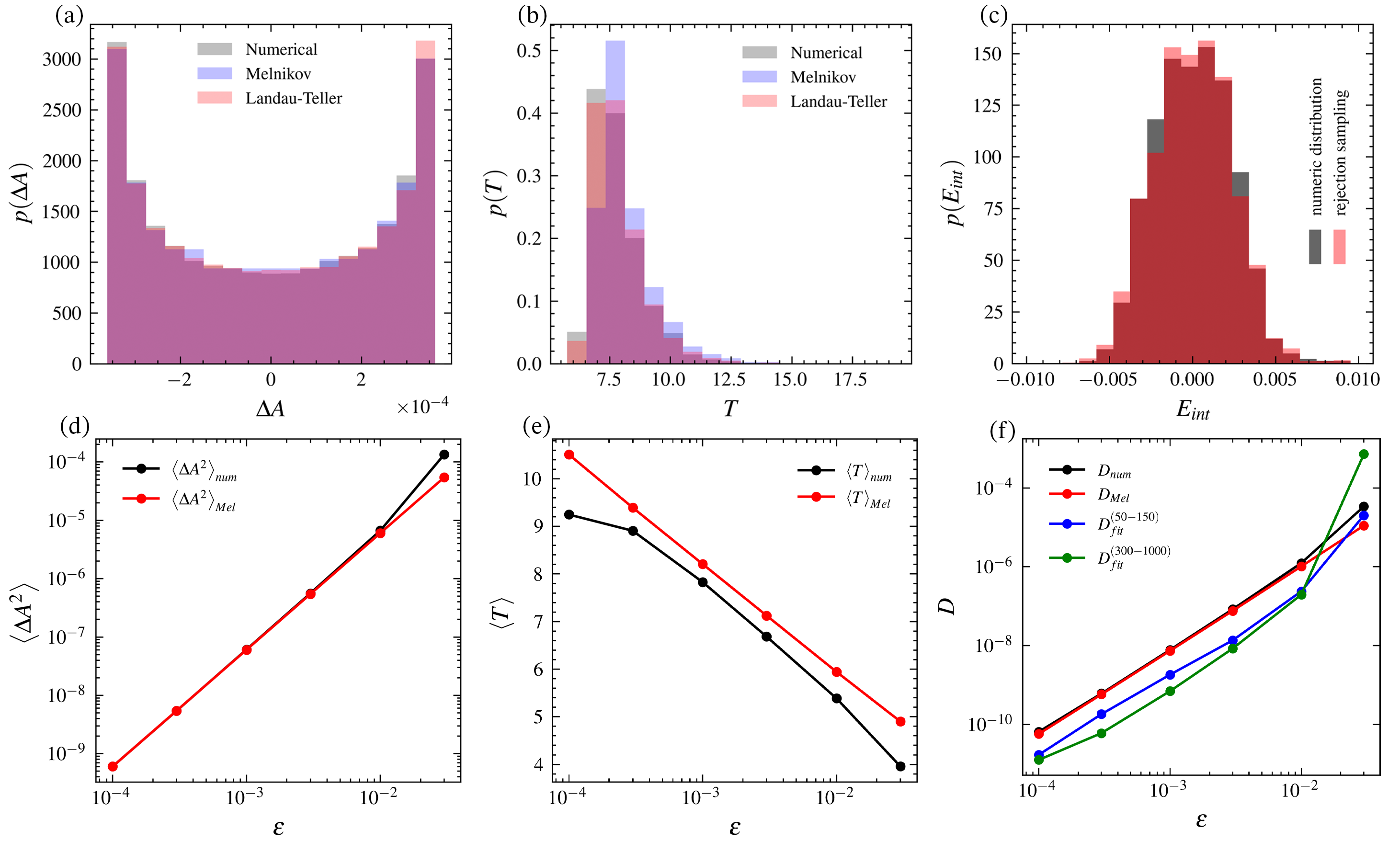}
    \caption{\small (a) Numerical (gray) histogram of the probability distribution $P(\Delta A)$ of the individual jumps in the value of the adiabatic action $A(t)$, collected by the data of the complete ensemble of simulated trajectories for $\varepsilon=0.001$, versus the corresponding histogram by the Melnikov approximation (magenta, Eq.(\ref{eq:probda})). 
    A third estimate of the same histogram (orange) is obtained independently by the stochastic Landau-Teller model of subsection \ref{sec:stochastic_LT}. 
    (b) Comparison of the three histograms obtained as in (a) for the PDF of the periods $p(T)$. (c) PDF of the values of the energies of the integrable model $E=H_{res}$ obtained at the middle of each homoclinic loop. 
    (d) The mean-square values $<\Delta A^2>$ (black) found by the numerical data and the corresponding Melnikov estimates $<\Delta A_M^2>$ (red) as a function of $\varepsilon$. 
    (e) The mean period $<T>$ by the numerical data (black) and the Melnikov estimate $<T_M>$ (subsection \ref{sec:Melnikov}) as a function of $\varepsilon$. 
    (f) The estimates on the diffusion coefficients with the assumptions of normal diffusion and uncorrelated phases $D=<\Delta A^2>/<T>$ for the numerical data and by the Melnikov approach, as a function of $\varepsilon$. The two lowermost curves in the same graph show the numerical diffusion coefficients as a function of $\varepsilon$, obtained by linear fit as in Fig.\ref{fig:diffusion} in the intervals $50\leq t\leq 150$ (blue) or $300\leq t\leq 1000$ (green) (see subsection \ref{sec:numprobes}).}
\label{fig:histo}
\end{figure}
Figure \ref{fig:histo}a shows that the probability distribution function (hereafter PDF) for the jumps $\Delta A$ as computed by collecting the values of all the jumps obtained numerically (by the simulated ensemble of trajectories, same as in Fig.\ref{fig:diffusion}), or semi-analytically (using formula (\ref{eq:A_melnikov01})), compare excellently. In fact, in the particular example dealt with, since $S_\infty(t)$, $\sigma_\infty(t)$ are odd and even functions of $t$ respectively, we readily find:
\begin{equation}\label{eq:daI1}
\Delta A_M(\phi_0)=\varepsilon\lambda\sin\phi_0 I_1
\end{equation}
where
$$
I_1= \int_{-\infty}^{\infty} \hat{S}_\infty^{2}(u) \cos \left(2\hat{\sigma}_\infty(u)-\Omega u\right) du~~.
$$
Hence, with $\phi_0$ uniform in $[0,2\pi)$, we readily find the theoretical estimate for the PDF of the jumps 
\begin{equation}\label{eq:probda}
    P(\Delta A_M)={1\over\pi}{1\over\sqrt{\varepsilon^2I_1^2-\Delta A_M^2}}~~.
\end{equation}
The PDF (\ref{eq:probda}) implies $<\Delta A_M>=0$ and 
\begin{equation}\label{eq:da2eps}
<\Delta A_M^2>=\varepsilon^2 I_1^2/2~.
\end{equation}

The good representation of the jumps by the Melnikov approximation is due to the fact that near the maximum of the homoclinic pulse, which corresponds to the values $\sigma_c=0$ or $\sigma_c=\pi$, we have that $\dot{\sigma}_c=\dot{\sigma} \rvert_{\sigma=0}$ and the frequencies $\Omega,\Omega_2$ are of comparable order of magnitude. As seen in Fig.\ref{fig:swarm} (compare panels c,e,and f), this implies that the pulse in $S(t)$ acts as an envelope that modulates the amplitude of the \textit{maximum one} complete sinusoidal oscillation of $\dot{A}$, which is accomplished along the homoclinic pulse. 
The jump then is produced by the difference in the areas under the positive and the negative part of the graph of $\dot{A}$ along the oscillation. 
In fact, as seen also in Fig.\ref{fig:swarm}f, the critical argument $2\sigma-\phi$ in $\dot{A}$ circulates with a period $T_c\simeq 1.3$, which is comparable to the half-width of the homoclinic pulse (Fig.\ref{fig:swarm}c). In the terminology of \cite{guzzo2020semi}, this is a case where the `quasi-stationary phase approximation' applies. Note, however, that, contrary to the discussion in \cite{guzzo2020semi}, in the present example (which is typical of models related to secular resonances in Celestial Mechanics), the quasi-stationarity is established at the level of the original Hamiltonian and requires no normal form analysis based on small divisors. 
On the other hand, we have checked numerically that, by choosing parameters $\Omega>>\dot{\sigma}_c$, we arrive at jumps whose size cannot be recovered by the Melnikov approximation unless the original Hamiltonian is processed and an optimal (with respect to the remainder) normal form model for $H_r$ is used, instead.

Table \ref{tab:num_vs_mel}(columns 1-3) summarizes the agreement between the semi-analytical and numerical estimates for $<\Delta A^2>$ obtained for different values of $\varepsilon$ in a range of 3.5 orders of magnitude. This information is depicted in Fig.\ref{fig:histo}b, which makes also evident the scaling $<\Delta A^2>\propto\varepsilon^2$, as predicted by the theory. 
\begin{table}
    \begin{center}
    \begin{adjustbox}{width=\textwidth}
    \begin{tabular}{ccccccc}
    \toprule
      $\varepsilon$  & $\left\langle \Delta A ^2 \right\rangle_{n}$   & $\left\langle \Delta A ^2 \right\rangle_{Mel}$  & $\left\langle T \right\rangle_{n}$ & $\left\langle T \right\rangle_{Mel}$  & $D_n$ & $D_{Mel}$\\
    \midrule
    $3 \times 10^{-2}$ & $1.34 \times 10^{-4}$  & $5.43 \times 10^{-5}$ & $3.96$ & $4.90$ & $3.39 \times 10^{-5}$  & $1.11 \times 10^{-5}$ \\ 
    $1 \times 10^{-2}$ & $6.67 \times 10^{-6}$  & $6.04 \times 10^{-6}$ & $5.39$ & $5.94$ & $1.24 \times 10^{-6}$  & $1.02 \times 10^{-6}$ \\ 
    $3 \times 10^{-3}$ & $5.60 \times 10^{-7}$  & $5.43 \times 10^{-7}$ & $6.16$ & $6.64$ & $8.37 \times 10^{-8}$  & $7.62 \times 10^{-8}$ \\ 
    $1 \times 10^{-3}$ & $6.11 \times 10^{-8}$  & $6.04 \times 10^{-8}$ & $7.83$ & $8.21$ & $7.80 \times 10^{-9}$  & $7.36 \times 10^{-9}$ \\ 
    $3 \times 10^{-4}$ & $5.41 \times 10^{-9}$  & $5.43 \times 10^{-9}$ & $8.91$ & $9.39$ & $6.08 \times 10^{-10}$  & $5.78 \times 10^{-10}$ \\ 
    $1 \times 10^{-5}$ & $6.05 \times 10^{-10}$ & $6.04 \times 10^{-10}$ & $9.25$ & $10.51$ & $6.54 \times 10^{-11}$  & $5.74 \times 10^{-11}$ \\ 
    \botrule
    \end{tabular}
    \end{adjustbox}
    \caption{\small Numerical values for different choices of $\varepsilon$ (with $\lambda = 1.2$) comparing the mean squared jump, mean period and diffusion coefficient $ < \Delta A ^2 >/ < T > $ for the numerical computation and the semi-analytic approach.
    }\label{tab:num_vs_mel}%
    \end{center}
    \end{table}

\subsection{Stochastic Landau-Teller model}
\label{sec:stochastic_LT}
\begin{figure}
    \centering
    \includegraphics[width=0.5\textwidth]{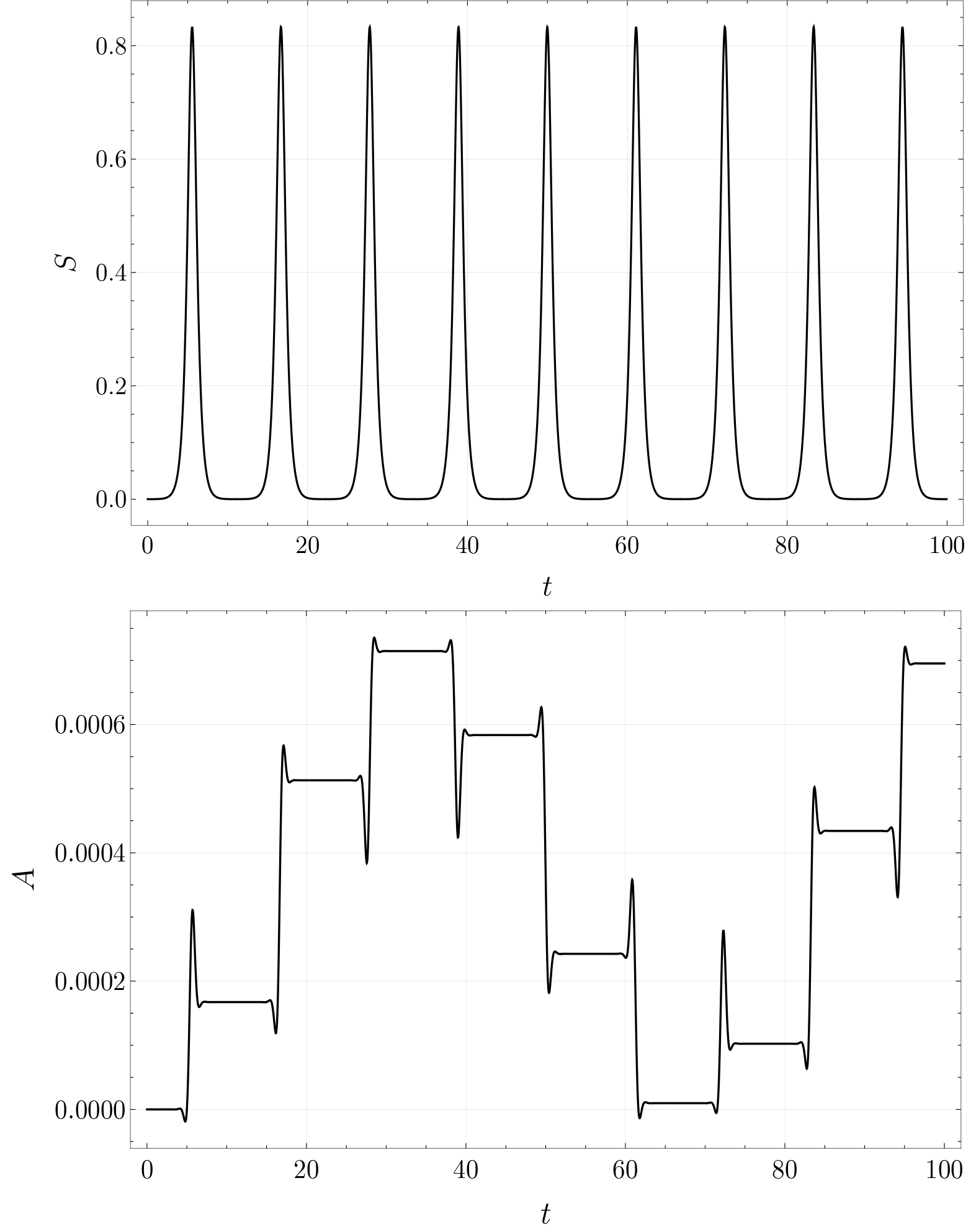}
    \caption{\small Top: evolution of $\hat{S}(t)$ for $\varepsilon=0.001$, $E=-5\times 10^{-5}$. Bottom: the curve $A(t)$ generated by the integral (\ref{eq:A_melnikov}) for the above solution. The graph of $A(t)$ exhibits jumps qualitatively similar as those of Fig.\ref{fig:swarm}d, and yielding a similar PDF as in Fig.\ref{fig:histo}a in the limit $t\rightarrow\infty$.}
   \label{fig:jumpslt}
\end{figure}

Consider again the Melnikov approximation (\ref{eq:A_melnikov}), in which a \textit{finite period} librational solution $\hat{S}(t),\hat{\sigma}(t)$ close to the separatrix ($E\lesssim 0$) is substituted in the integrand of the corresponding Melnikov integral. Let $T(E)$ be the period. The graphs of $\hat{S}(t)$ and $\hat{\sigma}(t)$ mimic the corresponding graphs of the numerical homoclinic pulses shown in Figs.\ref{fig:swarm}c,f, but the pulses are now repeated periodically, with period $T(E)$. The consecutive phases $\phi_0^{(j)}, \phi_{0,2}^{(j)}, j=1,2,\ldots$ differ by 
\begin{equation}\label{phaseslt}
    \phi_0^{(j)}=\phi_0^{(j-1)} +\Omega T(E)~,~~~
    \phi_{0,2}^{(j)}=\phi_{0,2}^{(j-1)} +\Omega_2 T(E)~~.
\end{equation} 
Taken modulo $2\pi$, these phases fill uniformly the interval $[0,2\pi)$ provided that the frequencies $\Omega,\Omega_2$ and $2\pi/T(E)$ are incommensurable. Also, for $E\lesssim 0$ we have $\hat{S}(t)\approx S_\infty(t), \hat{\sigma}(t)\approx \sigma_\infty(t)$ in the interval $-T(E)/2\leq t\leq T(E)/2$. It follows that, as $t\rightarrow\infty$, the Melnikov integral (\ref{eq:A_melnikov}) will produce a sequence of infinitely many jumps in the value of $A(t)$. Furthermore, while the homoclinic pulses of $\hat{S}(t)$ are repeated periodically, consecutive jumps do not repeat periodically since they correspond, in general, with different, albeit perfectly correlated (according to Eq.(\ref{phaseslt})) phases $\phi_0^{(j)}$ (see Fig.\ref{fig:jumpslt}). 

Two key remarks regarding the above are: 
 
(i) the PDF $P(\Delta A_{LT})$, called, hereafter, the `probability distribution function of the Landau-Teller jumps $\Delta A_{LT}$', practically coincides with the PDF of the Melnikov jumps $P(\Delta A_M)$. We readily find
\begin{equation}\label{eq:probdalt}
    P(\Delta A_{LT})={1\over\pi}{1\over\sqrt{\varepsilon^2I_{LT}^2-\Delta A_{LT}^2}}~~.
\end{equation}
where
$$
I_{LT}= \int_{-T(E)/2}^{T(E)/2} \lambda \hat{S}^{2}(u) \cos \left(2\hat{\sigma}(u)-\Omega u\right) du~~.
$$
The near-equality follows from the relation $I_1\simeq I_{LT}$, which, in turn, follows from the relations $\hat{S}_\infty\simeq 0$ if $t<-T(E)/2$ or $t>T(E)/2$, and $\hat{S}_\infty(t)\simeq\hat{S}(t)$ for $-T(E)/2\leq t\leq T(E)/2$. 

(ii) On the other hand, one can easily verify that, for any periodic solution $\hat{S}(t)$ and $\hat{\sigma}(t)$ substituted in the integral (\ref{eq:A_melnikov}), $A(t)$ (and, analogously, $A_2(t)$) is bounded at all times $t$, namely there can be no diffusion in the adiabatic action variables. To demonstrate this fact, consider the Fourier series representations:
\begin{eqnarray}\label{eq:fourierlt} 
\hat{S}^2(t)&=&\sum_{k=0}^{\infty} a_k \cos\left(k \frac{2 \pi}{T_S} t \right)
\nonumber\\
\sin(2\hat{\sigma}(t))&=&\sum_{k=1}^{\infty} s_k \sin\left(k \frac{2 \pi}{T_S} t \right)
\\
\cos(2\hat{\sigma}(t))&=&\sum_{k=0}^{\infty} c_k \cos\left(k \frac{2 \pi}{T_S} t \right)~~\nonumber
\end{eqnarray}
where $T_S=T(E)$ is the period of a librational periodic solution $(\hat{S}(t),\hat{\sigma}(t))$ of the integrable Hamiltonian $H_{res}$ with $E\lesssim 0$. The use of only sines or cosines in the above Fourier series follows from the parity of the functions $\hat{S}(t)$ (even) and $\hat{\sigma(t)}$ (odd). Set $\Omega_S=2\pi/T_S$. From Hamilton's equations, the following proposition immediately holds: \\
\\
\noindent
{\bf Proposition 1:} The Melnikov solutions
\begin{eqnarray}
    \label{eq:A_melnikov02}
    \hat{A}(t) &=&\hat{A}(0)-\int_{0}^{t} \varepsilon \lambda \hat{S}^{2}(u) \sin \left(2\hat{\sigma}(u)- (\phi(0)+\Omega u)\right) du \\
    \hat{A}_2(t) &=&\hat{A}_2(0)-\int_{0}^{t} \varepsilon \lambda \hat{S}^{2}(u) \sin \left(2\hat{\sigma}(u)- (\phi_2(0)+\Omega_2 u)\right) du \nonumber
\end{eqnarray}
satisfy Hamilton's equations for the canonical pairs $(\hat{A},\phi)$, $(\hat{A}_2,\phi_2)$, $(\hat{S},\hat{\sigma})$, under the `Landau-Teller' Hamiltonian
\begin{eqnarray}\label{eq:hamlt}
H_{LT}&=&\Omega\hat{A}+\Omega_2\hat{A}_2+\Omega_S \hat{A}_S  
\nonumber\\  
&+&\varepsilon\lambda
\left(\sum_{k=0}^{\infty} a_k \cos(k \phi_S )\right)
\Bigg[
\left(\sum_{\ell=0}^{\infty} c_\ell \cos(\ell \phi_S )\right)(\cos(\phi)+\cos(\phi_2))  
\nonumber\\
&+&
\left(\sum_{\ell=0}^{\infty} s_\ell \sin(\ell \phi_S )\right)(\sin(\phi)+\sin(\phi_2))  
\Bigg] 
\end{eqnarray}
Furthermore:\\
\\
\noindent
{\bf Proposition 2:} Assume the frequency vector $\mathbf{\Omega}\equiv(\Omega_S,\Omega,\Omega_2)$ satisfies the diophantine condition
\begin{equation}\label{eq:dioph}
\mid\mathbf{k}\cdot\mathbf{\Omega}\mid\geq{\gamma\over\mid\mathbf{k}\mid^{\tau}}
\end{equation}
for some positive constants $\gamma,\tau$, and for all $\mathbf{k}\equiv(k_S,k,k_2) \in \mathbb{Z}^3/0$.  Then, the Hamiltonian $H_{LT}$ is Arnold-Liouville integrable, and admits only bounded solutions for the action variables $\hat{A}$, $\hat{A}_2$. 

\begin{proof}
By the analyticity of the functions $\hat{S}(t),\hat{\sigma}(t)$, the Fourier coefficients $a_k,c_k,s_k$ decay exponentially, i.e., there exist positive constants $C_1,C_2,C_3$ and $d$ such that $\mid a_k\mid \leq C_1 e^{-d k}$, $\mid c_k\mid \leq C_2 e^{-d k}$, $\mid s_k\mid \leq C_3 e^{-d k}$, for all $k\in\mathbb{N}$.  Taking into account the even parity of $H_{LT}$ with respect to the change of sign $(\phi,\phi_2,\phi_S)\rightarrow(-\phi,-\phi_2,-\phi_S)$, the Hamiltonian $H_{LT}$ can be recast as
\begin{eqnarray}\label{eq:hamlt2}
    H_{LT}&=&\Omega\hat{A}+\Omega_2\hat{A}_2+\Omega_S \hat{A}_S  \\
    &+&\varepsilon\lambda
    \sum_{k_S=0}^{\infty}
    \sum_{k=-\infty}^{\infty}
    \sum_{k_2=-\infty}^{\infty}
     d_{k_S,k,k_2} \cos(k_S\phi_S +k\phi+k_2\phi_2)  \nonumber
\end{eqnarray}
where the coefficients $d_{k_S,k,k_2}$ also decay exponentially, $\mid d_{k_S,k,k_2}\mid\leq C_1Ce^{s(\mid\mathbf{k}\mid)}$, with $C=\max(C_2,C_3)$. Consider now the formal canonical change of variables:
\begin{eqnarray}\label{eq:lietra} 
    \phi_S'&=&\phi_S \nonumber\\
    \phi'&=&\phi     \nonumber\\
    \phi_2'&=&\phi_2              \\
    \hat{A}_S'&=&\hat{A}_S+\varepsilon\lambda
    \sum_{k_S=1}^{\infty}
    \sum_{k=-\infty}^{\infty}
    \sum_{k_2=-\infty}^{\infty}
     {k_Sd_{k_S,k,k_2}\cos(k_S\phi_S +k\phi+k_2\phi_2)
     \over{k_S\Omega_S+k\Omega+k_2\Omega_2}}
    \nonumber\\
     \hat{A}'&=&\hat{A}+\varepsilon\lambda
     \sum_{k_S=0}^{\infty}
     \sum_{k=-\infty,k!=0}^{\infty}
     \sum_{k_2=-\infty}^{\infty}
     {kd_{k_S,k,k_2}\cos(k_S\phi_S +k\phi+k_2\phi_2)
     \over{k_S\Omega_S+k\Omega+k_2\Omega_2}}
    \nonumber\\
      \hat{A}_2'&=&\hat{A}_2+\varepsilon\lambda
      \sum_{k_S=0}^{\infty}
      \sum_{k=-\infty}^{\infty}
      \sum_{k_2=-\infty,k_2!=0}^{\infty}
      {k_2d_{k_S,k,k_2}\cos(k_S\phi_S +k\phi+k_2\phi_2)
      \over{k_S\Omega_S+k\Omega+k_2\Omega_2}}
    \nonumber
 \end{eqnarray}   
In view of the diophantine condition (\ref{eq:dioph}) and the exponential bound on the coefficients $d_\mathbf{k}\equiv d_{k_S,k,k_2}$, we have that
\begin{equation}\label{eq:cfbound}
{\mid\mathbf{k}\mid \mid d_\mathbf{k}\mid
\over\mid\mathbf{k}\cdot\mathbf{\Omega}\mid}
\leq
{\mid\mathbf{k}\mid^{\tau+1}{C_1C\over\gamma}e^{-s\mid\mathbf{k}\mid}}
\leq
\left({2(\tau+1)\over es}\right)^{\tau+1}
\left({C_1C\over\gamma}\right)e^{-s\mid\mathbf{k}\mid/2}~~.
\end{equation}   
Hence, all the Fourier sums in (\ref{eq:lietra}) are convergent, and we obtain the explicit bounds $\mid\hat{A}'-\hat{A}\mid\leq\delta A$, $\mid\hat{A}_2'-\hat{A}_2\mid\leq\delta A$, $\mid\hat{A}_S'-\hat{A}_S\mid\leq\delta A$, with 
\begin{equation}
    \delta A = 8\pi \varepsilon\lambda {1\over\gamma}s^{-(\tau+4)}\Gamma(\tau+4)
\end{equation}
However, substituting the transformation (\ref{eq:lietra}) into the Hamiltonian, we get, in the new variables:
\begin{equation}
    H_{LT}=\Omega\hat{A}'+\Omega_2\hat{A}_2'+\Omega_S \hat{A}_S'~~.
\end{equation}

Thus, all three new actions $\hat{A}',\hat{A}_2',\hat{A}_S'$ are integrals of motion. It follows that the Hamiltonian $H_{LT}$ is Arnold-Liouville integrable, and for any trajectory the actions $\hat{A},\hat{A}_2$ are bounded to a zone of width $\delta A$ around the corresponding (constant) values $\hat{A}',\hat{A}_2'$ for the same trajectory. 
\end{proof}

Figure \ref{fig:jumpslt} illustrates the results of the previous analysis. The evolution of $\hat{A}(t)$ under the Landau-Teller Hamiltonian is shown together with the solution $\hat{S}(t)$ for $E=-5\times 10^{-5}$. Note the similarity between the homoclinic pulses in the Landau-Teller model (Fig.\ref{fig:jumpslt}a) and in the complete 3DOF Hamiltonian (Fig.\ref{fig:swarm}c), whose only essential difference is the lack of exact periodicity in the latter case. Note also the production of jumps in $\hat{A}(t)$, one per homoclinic pulse (Fig.\ref{fig:jumpslt}b), which, however, in the Landau-Teller evolution lead to no long-term diffusion as in the full 3DOF system, since they are bounded by the estimates of proposition 2 above. 

\begin{figure}
    \centering
    \includegraphics[width=1.\textwidth]{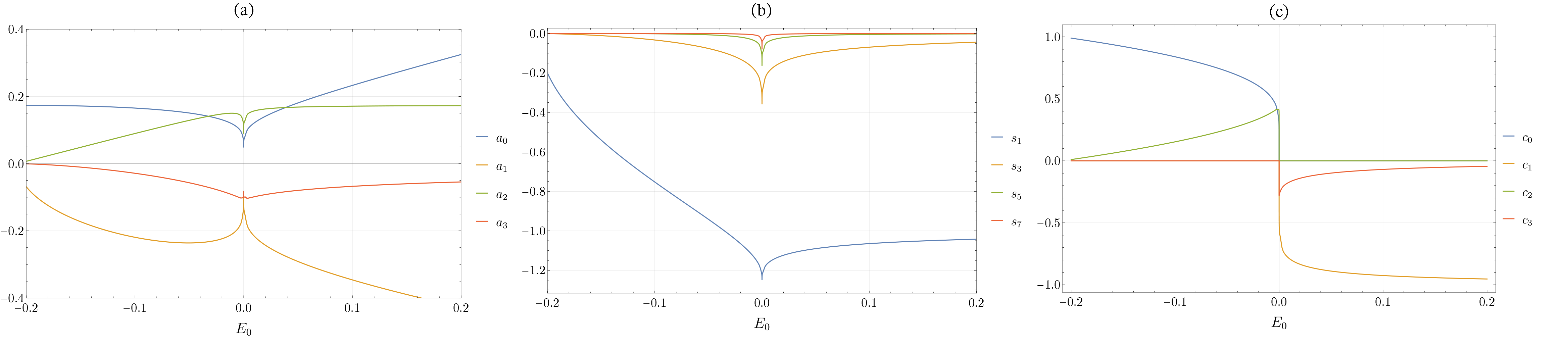}
    \caption{\small The first few coefficients $a_k$ (a), $c_k$ (b) and $s_k$ (c), $k=0,\ldots,4$, of the Fourier series (\ref{eq:fourierlt}) as a function of the energy $E$ which labels the corresponding homoclinic pulse.}
\end{figure}
Chirikov's whisker (or `separatrix') map \cite{chirikov1979universal} breaks the periodicity of the homoclinic pulses by introducing a deterministic, but chaotic, map, computed via Melnikov's sub-harmonic integral, which yields a sequence of values $E^{(j)}$, $j=1,2,\ldots$, each corresponding to the energy $E$ labelling the $j-$th pulse. In the limit of strong chaos, the sequence $E^{(j)}$ effectively becomes a sequence of randomly chosen values obtained from a PDF $p(E)$. Figure \ref{fig:histo}c shows the distribution $p(E)$ as found from the entire data set of simulated trajectories under the complete 3DoF model for $\varepsilon=0.001$, from which the PDF $p(T)$ of the associated periods (of the pulses) can be derived (Fig.\ref{fig:histo}b). Note that these PDFs can be approximated semi-analytically assuming ergodicity within the figure-8 shaped stochastic layers of Fig.\ref{fig:pmap3D}, and estimating their widths via Melnikov's sub-harmonic integral (see next section). At any rate, departing from given PDFs $p(E)$ (and, hence, $p(T)$), and using the functions $a_k(E)$, $c_k(E),s_k(E)$, a \textit{stochastic Landau-Teller Hamiltonian model} can be introduced, in which to compute the evolution of the adiabatic actions $A(t),A_2(t)$: this consists in computing the solutions to the equations of motion under the stochastic Landau-Teller Hamiltonian:
\begin{eqnarray}\label{eq:hamltsto}
    H_{LTS}&=&\Omega\hat{A}+\Omega_2\hat{A}_2+{2\pi\over T(E^{(j)})} \hat{A}_S  
    \nonumber\\  
    &+&\varepsilon\lambda
    \left(\sum_{k=0}^{\infty} a_k(E_j) \cos(k \phi_S )\right)
    \Bigg[
    \left(\sum_{\ell=0}^{\infty} c_\ell(E_j) \cos(\ell \phi_S )\right)(\cos(\phi)+\cos(\phi_2))  
    \nonumber\\
    &+&
    \left(\sum_{\ell=0}^{\infty} s_\ell(E_j) \sin(\ell \phi_S )\right)(\sin(\phi)+\sin(\phi_2))  
    \Bigg]~~~,\nonumber\\
    &~&\mbox{in the interval~}
    \sum_{n=1}^{j-1} T(E^{(n)})\leq t\leq \sum_{n=1}^{j} T(E^{(n)}),~~~j=1,2,\ldots \nonumber 
\end{eqnarray}
Since the Hamiltonian $H_{LTS}$ is piecewise integrable (for all times $t$ within the same $j-$th interval), the evolution of the adiabatic actions $\hat{A}(t)$ and $\hat{A}_2(t)$ can be computed analytically for any random sequence $E^{(j)}$ obeying the adopted PDF $p(E)$. Figure \ref{fig:histo}a shows, on top of the previously discussed histograms, the one of the jumps $\Delta A_{LT}$ produced by just \textit{one} trajectory integrated in the stochastic model $H_{LTS}$, after numerically defining (via the `rejection algorithm', see \cite{press2007numerical}) a random sequence of values $E^{(j)}$ obeying the statistics of Fig.(\ref{fig:histo}c), and truncating all Fourier series at Fourier order $K=10$. We note that all three histograms, i.e. (i) fully numerical, (ii) obtained through the random phase approximation of the Melnikov integrals, or (iii) through the stochastic Landau-Teller model, compare excellently to each other.  

\subsection{Phase correlations and estimates on the speed of diffusion}
Consider an ensemble of $N$ trajectories, and let $A_i(t)$ be the evolution of the adiabatic action $A$ for the i-th trajectory. Assume that, modulo small ($O(\varepsilon)$) oscillations, the evolution of $A_i$ proceeds by consecutive discrete jumps $\Delta A_i^{(j)}$. If the mean period of homoclinic pulses is $<T>$, there are $n=[t/<T>]$ jumps taking place per each trajectory. Setting the initial condition $A_i(0)=0$ for all $i=1,\ldots,N$, we have
\begin{equation}\label{eq:sumda}
A_i(t = n <T>)=\sum_{j=1}^n\Delta A_i^{(j)}~~~.
\end{equation}
Assume now that the sequence of jumps $\Delta A_i^{(j)}$ exhibits no correlation, i.e., the individual jumps $\Delta A_i^{(j)}$ are randomly chosen from a distribution $p(\Delta A)$ with mean $\mu=0$ and dispersion $\sigma_{\Delta A}$. By the central limit theorem, for $n$ large, we then have
\begin{equation}\label{eq:normal}
A_i\sim{\cal N}(0,n\sigma_{\Delta A}^2)~~~.
\end{equation}
Since $\mu=0$, we have $\sigma_{\Delta A}^2=<\Delta A^2>$. Hence, the linear diffusion coefficient for the ensemble satisfies the relation:
\begin{equation}\label{eq:difda2t}
D={n\sigma_{\Delta A}^2\over n<T>}={\Delta A^2\over<T>}
\end{equation}
Figure \ref{fig:histo}f compares the estimate (\ref{eq:difda2t}) with the quasi-linear diffusion coefficients obtained by direct fitting on the numerical data (as in Fig.\ref{fig:diffusion}), as a function of $\varepsilon$, varied within three orders of magnitude, from $\varepsilon=10^{-4}$ to $\varepsilon=10^{-1}$. Note that, since the numerical histograms found for the PDF $p(\Delta A)$ practically coincide with those found either by the Melnikov or Landau-Teller estimates, the corresponding curves $D(\varepsilon)$ practically coincide as well, small differences being only due to the estimates for $<T>$ (Fig.\ref{fig:histo}e). In particular, by Eq.(\ref{eq:da2eps}), we have $<\Delta A^2>\propto\varepsilon^2$, while $<T>$ has a weak power-law dependence on $\varepsilon$, $<T>\propto\varepsilon^{-\alpha}$ with $\alpha\approx 0.13$. Thus, the estimate (\ref{eq:difda2t})) yields $D\sim\varepsilon^{2+\alpha}\approx\varepsilon^{2.13}$. As shown in Fig.\ref{fig:histo}f, this power-law dependence of the quasi-linear diffusion coefficient on $\varepsilon$ is reproduced by the numerical fitting data, which, however, yield numerical values of the diffusion coefficient systematically lower (by about one order of magnitude, for all values of $\varepsilon$ except for $\varepsilon=0.1$) than the estimate (\ref{eq:difda2t}). This tendency is systematically found with various choices of models, and also observed in the case of diffusion models for the navigation satellites (see next section), it indicates that the assumption of effectively completely uncorrelated jumps $\Delta A^{(j)}$ (i.e., phases $\phi^{(i)})$ is not fulfilled by the numerical data. In Chirikov's terminology (see )\cite{chirikov1979universal}), phase correlations result in the estimate $D\sim F<\Delta A>^2/<T>$, where $F$ is called the `reduction factor'. Chirikov suggests $F\approx 1/3$ in the case of the first fundamental model of resonance, while our data, here and in the next section as well, suggest $F\approx 0.1$ far from the strong chaos regime.

\section{Diffusion along the $2g+h$ resonance}
\label{sec: diffusion_2g+h}
\label{sec: diffusion_2g+h}
We now apply the theories illustrated so far in order to provide a semi-analytical characterization of the diffusion properties of navigation satellites.

We call a satellite navigation system with global coverage a \textit{global navigation satellite system} (GNSS).
These satellites are located in the MEO region\footnote{
    A medium Earth orbit (MEO) is an Earth-centered orbit with an altitude between $2000$ and $35786$ km above sea level, i.e. above a low Earth orbit (LEO) and below a geosynchronous orbit (GEO).}, 
where is present a web of lunisolar resonances that shape the dynamics (see \cite{Legnaro_Efthymiopoulos_2022}).
Lunisolar resonances have the effect of pumping an object's eccentricity, possibly up to a value high enough that the orbit's perigee reaches the atmosphere and then friction will cause the re-entering of the object (see \cite{rosengren2015chaos}, \cite{daquin2016dynamical}, \cite{Gondelach_Armellin_Wittig_2019}).

Currently, four global systems are operational inside the MEO region, and three of these are within the domain of the $2g+h$ resonance: the European GALILEO ($a \sim 29600$ km, $i \sim 55^\circ$), the American GPS ($a \sim 26560$ km, $i \sim 55^\circ$) and some of China's BEIDOU  ($a \sim 27906$ km, $i \sim 55^\circ$).

\noindent
For this reason, we will focus on the $2g + h$ resonance. 
The starting point will be the results presented in \cite{daquin2021deep} and \cite{Legnaro_Efthymiopoulos_2022}. 
We briefly recall here some necessary information.
Let $H (X,Y,J_F,u_F,A,\Omega_L)$ (Eq. 6 of \cite{Legnaro_Efthymiopoulos_2022}) be the resonant Hamiltonian for the $2g+h$ resonance. 
It has the structure $H=H_R+H_{CM}+H_M$
with $H_{CM}=H_{CM}^0+H_{CM}^1$ and
\begin{align}
\begin{split}
    &H_R=R_{20} X^{2}+R_{02} Y^{2}+R_{22} X^{2} Y^{2}+R_{40} X^{4}+R_{04} Y^{4}+\ldots \\
&H_{C M}^{0} =\omega_F J_{F}+\alpha J_{F}^{2}+C_1 \cos u_{F}+C_2 \cos 2 u_{F}+\ldots \\
&H_{C M}^{1} = A \, \nu_L+D_{21} \cos \left(2 u_{F}+\Omega_L \right)+D_{11} \cos \left(u_{F}+\Omega_L \right)+\ldots \\
&H_M = M_{120} J_{F} X^{2}+M_{102} J_{F} Y^{2}+\ldots
\end{split}
\label{eq: H_ILSR}
\end{align}
The indices "$R$" and "$CM$" in the above notation stand for \index{Center Manifold}\emph{Resonant} and \emph{Center Manifold} respectively while the terms in $H_M$ represent a coupling between resonant $(X, Y)$ and "Center Manifold" variables $(u_F, J_F, \Omega_L, A)$. 

\noindent
In the case of the $2g+h$ resonance, the fast action $J_F$ can be expressed in Keplerian elements as
\begin{equation}
    J_F(a, e, i)=\sqrt{\mu_E a}\left(\frac{1}{2}+\sqrt{1-e^2}\left(\cos i_{\star}-\cos i-\frac{1}{2}\right)\right).
\end{equation}
We define \textit{fast drift plane} (FDP) the locus of points where the fast action $J_F$ is constant.
Each of such planes is associated with an integrable phase portrait. This shows the figure-8 separatrix of the $2g+h$ resonance, and its extension dictates the theoretical maximum eccentricity value that an object would reach moving along the resonance.
Because of this, we are interested in studying the diffusion along fast drift planes.

By averaging over the angles $u_F$ and $\Omega_L$ of the resonant Hamiltonian and keeping just the leading terms, we get the integrable Hamiltonian
\begin{align}
\begin{split}
    H_{0}(&X, Y; a, J_F) =  \; \omega_F J_F + \alpha^2 J_F^2+R_{20} X^2+ \\
    & R_{02} Y^2+R_{40} X^4+ R_{04} Y^4 + M_{140} J_F X^4+ M_{104} J_F Y^4 +  \\
	& R_{60} X^6 + R_{06} Y^6+M_{120} J_F X^2+M_{102} J_F Y^2+ \\
	& M_{220}J_F^2 X^2 + M_{202} J_F^2 Y^2 + \\
	& R_{22} X^2 Y^2 + R_{42} X^4 Y^2 + R_{24} X^2 Y^4,
\end{split}
\label{eq: H_BR}
\end{align}
(see Appendix C of \cite{Legnaro_Efthymiopoulos_2022} for the coefficients' values).

\begin{figure}
    \centering
    \includegraphics[width = 0.7 \textwidth]{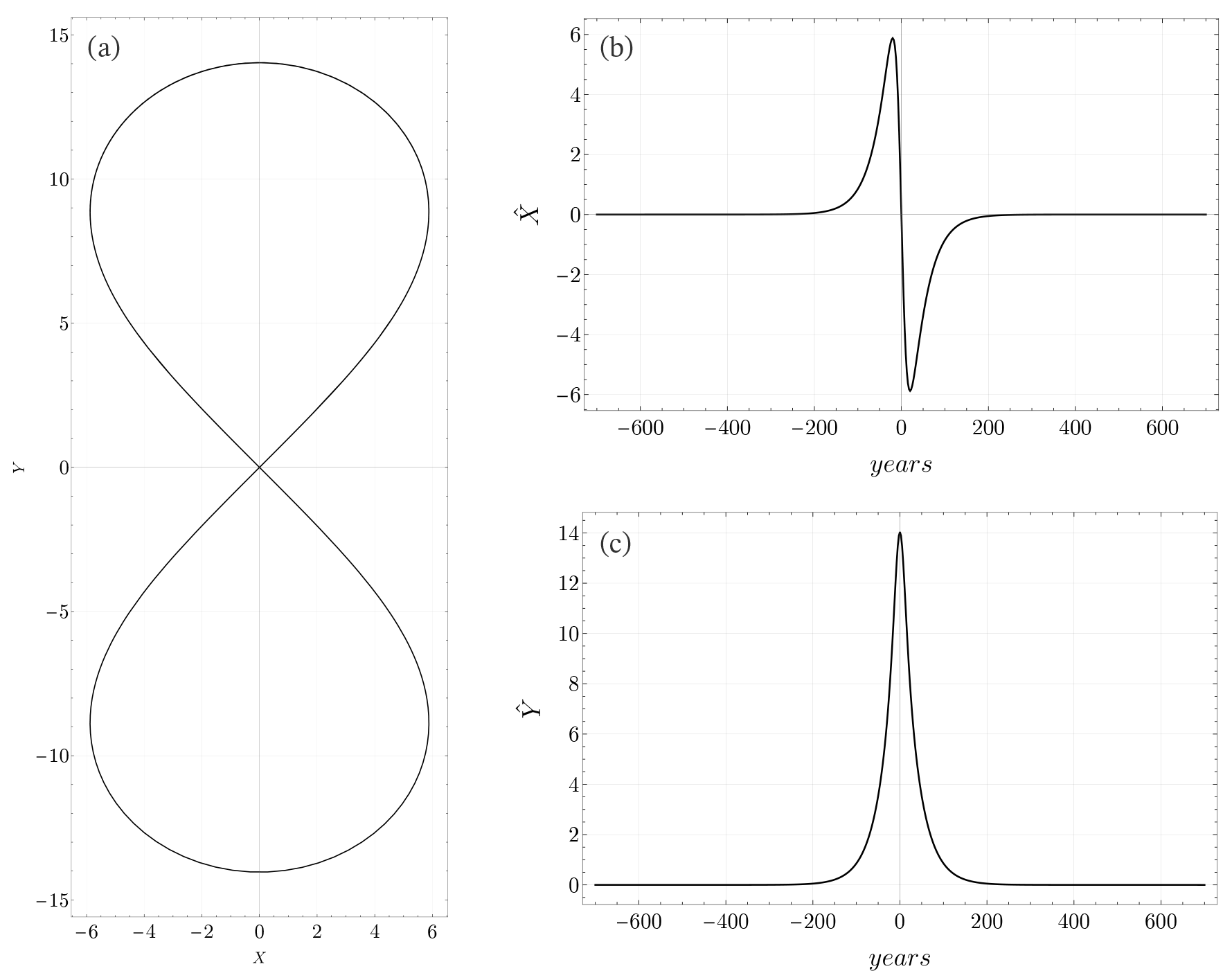}
    \caption{The integrable solutions of the separatrix (shown in the $X-Y$ plane (a)), $\hat{X}(t)$ (b) and $\hat{Y(t)}$ (c) for the Hamiltonian  \eqref{eq: H_BR} at the Galileo altitude $29600$ km.}
    \label{fig: Integrable_Hamiltonian_Galileo}
\end{figure}

\begin{figure}
    \centering
    \includegraphics[width = \textwidth]{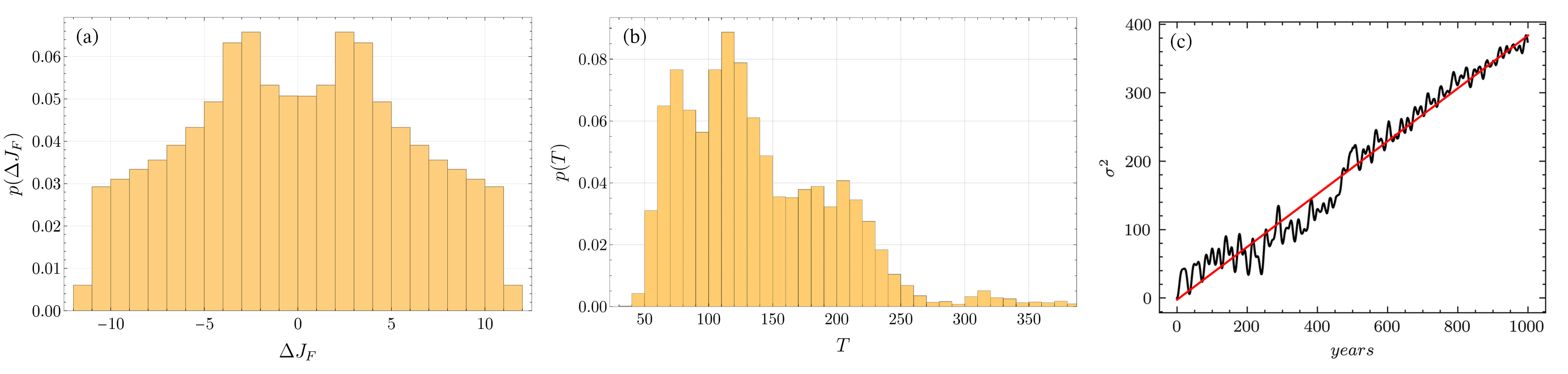}
    \caption{The histogram of jumps in $\Delta J_F$ (a), of the integrable periods (b) and the fitting of the numerical computation of $\sigma^2 (J_F)$ (c) up to $1000$ years for a Galileo-like satellite having $a = 29600$ km.}
    \label{fig: Galileo_Melnikov_Fitting}
\end{figure}

Let $\hat{X}(t)$ and $\hat{Y(t)}$ be the analytic solutions of the separatrix in the above Hamiltonian (see Figure \ref{fig: Integrable_Hamiltonian_Galileo}).
Going back to the resonant Hamiltonian $H$, define the Hamiltonian $\mathcal{H}$ as
\begin{equation}
    \mathcal{H} = H - H_{CM} + (\omega_F J_F + A \, \nu_L).
\end{equation}
The neglected terms only account for period oscillations around the resonance's center. 
Adopting the Melnikov approach, it is possible to derive a solution for the jump $\Delta J_F$ by integrating
\begin{equation}
    \Delta J_F = \int_{-\infty}^{\infty} - \frac{ \partial \mathcal{H}}{\partial u_F}(X, Y, u_F, \Omega_L) dt,
\end{equation}
where $X$ and $Y$ are substituted with the analytic solutions $\hat{X}(t)$ and $\hat{Y}(t)$ and the angles $u_F$, $\Omega_L$ evolve linearly with frequencies $\omega_F,\nu_L$ and initial data $u_F^0$, $\Omega_L^0$ respectively.
By doing so, we get a semi-analytic solution for the jump in $J_F$, and it is then possible to compute numerically the value of $\left< \Delta J_F^2 \right>$ (see the histogram for $\Delta J_F$ for Galileo altitudes in Fig. \ref{fig: Galileo_Melnikov_Fitting} (left)).

\begin{figure}[b]
    \centering
    \includegraphics[width = 0.9 \textwidth]{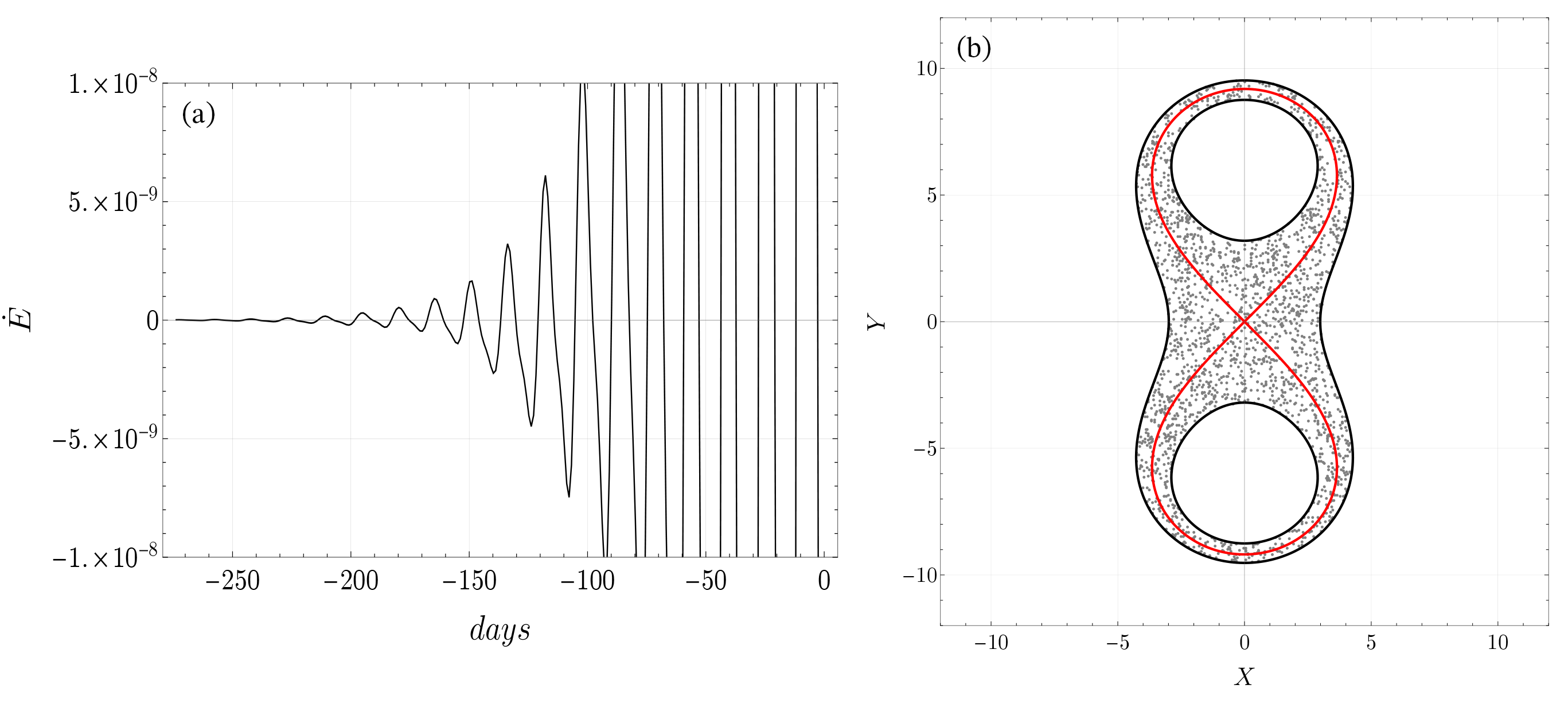}
    \caption{(a) The time evolution of \eqref{eq: Edot} for $a = 23000$ km.
    (b) Phase portrait at the same altitude of a uniform distribution of points inside the two contour levels corresponding to $\hat{E} - \Delta E$ and $\hat{E} + \Delta E$.}
    \label{fig: I_SubHarm_Meln_Integr}
\end{figure}

\begin{figure}
    \centering
    \includegraphics[width = \textwidth]{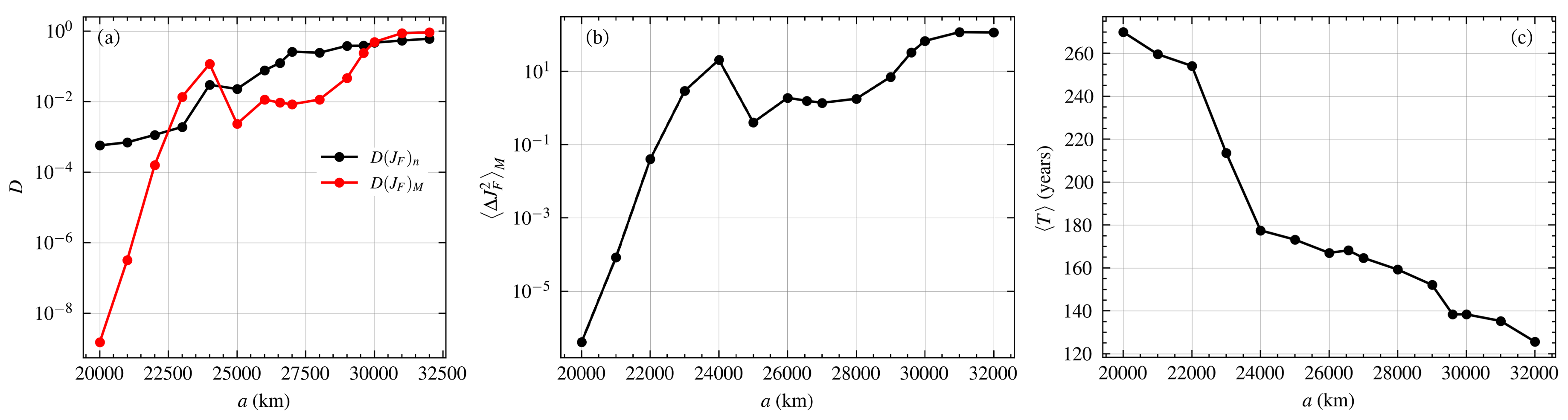}
    \caption{(a) Comparison between the semi-analytical diffusion coefficient $D(J_F)_M$ (red) and the numerically computed one $D(J_F)_n$ (black). (b) The value of $\left< \Delta J_F^2 \right>$ derived using the Melnikov approach. (c) The value of $\left< T \right>$ estimated computing the first subharmonic Melnikov integral. The numerical values are collected in Table \ref{tab: numerical values}.}
    \label{fig: plot_table_data}
\end{figure}

We can estimate the average period $\left< T \right>$ using the first subharmonic Melnikov integral, which in essence computes the displacement between the stable and unstable manifolds of the hyperbolic point, thus allowing an estimation of the chaotic layer's width in the $X,Y$ plane.
We first set $J_F = 0$.
Define $H_1 = \mathcal{H} - H_0$.
Let $t_1, t_2, \dots$ be the times at which 
\begin{equation}
    \dot{E} =  \left\lbrace H_0, H_1 \right\rbrace \biggr\rvert_{X = \hat{X}(t), Y = \hat{Y}(t), u_F = u_F^0 + t \omega_F, \Omega_L = \Omega_L^0 + \nu_L t }
    \label{eq: Edot}
\end{equation}
is equal to zero.
The time series $\dot{E}(t)$ of \eqref{eq: Edot} is shown in Fig. \ref{fig: I_SubHarm_Meln_Integr} (left).
Consider the maximum of $\Delta E = \max_i \Delta E_i$ where
\begin{equation}
    \Delta E_i = \int_{-\infty}^{t_i} \dot{E} dt.
\end{equation}
If $\hat{E}$ is the energy value of the separatrix, then the extension of the chaotic layer is approximated by the area with energy between $\hat{E} - \Delta E$ and $\hat{E} + \Delta E$.
Now, it is possible to estimate $\left< T \right>$ by generating a uniform distribution of points within these energy levels and then computing the integrable period for each point (see Figs. \ref{fig: I_SubHarm_Meln_Integr} and \ref{fig: Galileo_Melnikov_Fitting} for the histogram of the integrable periods at Galileo altitudes).
This allows to obtain semi-analytically a diffusion coefficient 
\begin{equation}
    D(J_F)_{M} \coloneqq \left< \Delta J_F^2 \right>/\left< T \right>
    \label{eq: D(J_F)_{M}}
\end{equation}
which can be compared with the numerically computed one $D(J_F)_n$ found by fitting the $\sigma^2 (J_F)$ curve (Fig. \ref{fig: Galileo_Melnikov_Fitting}c).
The numerical values for semi-major axis ranging from $20000$ to $32000$ km are reported in Table \ref{tab: numerical values} and plotted in Figure \ref{fig: plot_table_data}.

The semi-analytical prediction works best for higher altitudes. 
This is to be expected, since lower altitudes would require additional normalization steps, while at higher altitudes no additional perturbation step would make the remainder smaller and so the original Hamiltonian is already the optional one.

\begin{table}[b]
    \centering
    \begin{tabular}{ccccc}
        \toprule
    $a$ & $\left< \Delta J_F^2 \right>$ & $\left< T \right>$ & $D(J_F)_M$ & $D(J_F)_n$ \\
    \midrule
    $ 20000$ & $4.06 \times 10^{-7}$ & $ 269.92$ & $1.50 \times 10^{-9}$ & $5.73 \times 10^{-4}$ \\
    $ 21000$ & $8.33 \times 10^{-5}$ & $ 259.62$ & $3.21 \times 10^{-7}$ & $7.01 \times 10^{-4}$ \\
    $ 22000$ & $4.02 \times 10^{-2}$ & $ 254.15$ & $1.58 \times 10^{-4}$ & $1.14 \times 10^{-3}$ \\
    $ 23000$ & $ 2.93$ & $ 213.60$ & $1.37 \times 10^{-2}$ & $1.92 \times 10^{-3}$ \\
    $ 24000$ & $ 20.80$ & $ 177.49$ & $1.17 \times 10^{-1}$ & $3.00 \times 10^{-2}$ \\
    $ 25000$ & $ 0.41$ & $ 173.20$ & $2.34 \times 10^{-3}$ & $2.29 \times 10^{-2}$ \\
    $ 26000$ & $ 1.90$ & $ 167.01$ & $1.14 \times 10^{-2}$ & $7.74 \times 10^{-2}$ \\
    $ 26560$ & $ 1.58$ & $ 168.18$ & $9.39 \times 10^{-3}$ & $1.25 \times 10^{-1}$ \\
    $ 27000$ & $ 1.39$ & $ 164.67$ & $8.44 \times 10^{-3}$ & $2.62 \times 10^{-1}$ \\
    $ 28000$ & $ 1.79$ & $ 159.23$ & $1.15 \times 10^{-2}$ & $2.44 \times 10^{-1}$ \\
    $ 29000$ & $ 7.11$ & $ 152.11$ & $4.68 \times 10^{-2}$ & $3.83 \times 10^{-1}$ \\
    $ 29600$ & $ 32.95$ & $ 138.39$ & $2.38 \times 10^{-1}$ & $3.86 \times 10^{-1}$ \\
    $ 30000$ & $ 68.22$ & $ 138.36$ & $4.93 \times 10^{-1}$ & $4.72 \times 10^{-1}$ \\
    $ 31000$ & $ 118.11$ & $ 135.29$ & $8.73 \times 10^{-1}$ & $5.41 \times 10^{-1}$ \\
    $ 32000$ & $ 115.90$ & $ 125.59$ & $9.23 \times 10^{-1}$ & $6.09 \times 10^{-1}$\\
    \bottomrule
    \end{tabular}
    \caption{Numerical values of $\left< \Delta J_F^2 \right>$ derived using the Melnikov approach, $\left< T \right>$ estimated computing the first subharmonic Melnikov integral, their ratio \eqref{eq: D(J_F)_{M}} and the numerical diffusion coefficient $D(J_F)_n$ obtained by fitting with a line the curve $\sigma^2(J_F)$ on an interval that dependes on the altitude.}
    \label{tab: numerical values}
    \end{table}

\clearpage

\section{Conclusions}
\label{sec:conclusions}
In this paper, we first consider an archetype model representative of the dynamics in the case of secular resonances in Celestial Mechanics belonging to the SFM2 fundamental model of resonance. Our purpose is to find suitable semi-analytical models allowing to obtain i) the probability distribution function $P(\Delta A)$ of the jumps $\Delta A$ exhibited by orbits undergoing slow diffusion in the space of the adiabatic action variables $A$ of the system, and ii) semi-analytical estimates on the value of the diffusion coefficient under the assumptions of a locally normal character of the diffusion and of no correlations in the phase variables related to the process (see Section \ref{sec:theory}). 
Then, we applied the methods used in the analysis of the archetype model in a real case of secular resonance, namely the lunisolar resonance $2g+h$ in the problem of the motion of Earth's navigation satellites. Our main conclusions are: 

1) Using the solution $(\hat{S}_\infty(t),\hat{\sigma}_\infty(t))$ corresponding to an (infinitely long in time) homoclinic pulse along the separatrix of the integrable SFM model, in conjunction with the assumption of uniform distribution of the phases, we reach to reproduce semi-analytically the PDF $P(\Delta)$ observed in numerical experiments of diffusion involving a large ensemble of numerically computed trajectories (see subsection \ref{sec:Melnikov}). 
This is called a `Melnikov approach' to the computation of $P(\Delta A)$.

2) Alternatively, using a finite truncation of the Fourier representation of finite-time $T(E)$ homoclinic pulses, where $E$ is the energy of the integrable SFM Hamiltonian along a finite-time homoclinic loop, together with a stochastic process based on random values of $E$ obtained through a suitably defined distribution, we reach to have an alternative representation of $P(\Delta A)$, called the `stochastic Landau-Teller model'. 

3) Both the Melnikov and Landau-Teller approaches yield very precise local models of the jumps in the action variables along homoclinic loops, as well as equally precise estimates of the true (numerically observed) PDF $P(\Delta A)$. However, both models fail to account for long-term estimates on the speed and character (normal or not) of the chaotic diffusion in the space of the adiabatic action variables.
We attribute the failure to the existence of long-term phase correlations as anticipated already in \cite{chirikov1979universal} for the dynamical behavior of chaotic whisker maps. Nevertheless, order-of-magnitude precise estimates on the local values of (quasi-linear) diffusion coefficients can still be found by our semi-analytical models using a value of Chirikov's reduction factor $R\approx 0.1$. 

4) Implementing the above semi-analytical models to the case of navigation satellites, we arrive at order-of-magnitude correct estimates of the local diffusion coefficient along the separatrices of the $2g+h$ lunisolar resonances for all altitudes (values of the satellite's semi-major axis) except: 
a) close to major double-resonance points, and b) for semimajor axes $a<23000$ km. 
In the former case, the model of resonance can no longer be the SFM2. In the latter, the secular Hamiltonian found by averaging over all the fast angles of the problem is far from optimal, and we conjecture that the methods presented in this paper could still be applicable after obtaining an optimal resonant normal form model, i.e., one in which the function $H_{1,a}$ (Eq. \eqref{eq:hamgen}) represents a remainder minimized with respect to all possible choices of order of normalization. 
jumps. 

\bmhead{Acknowledgments}
We acknowledge the support of the Marie Curie Initial Training Network Stardust-R, grant agreement
Number 813644 under the H2020 research and innovation program: https://doi.org/10.3030/813644. 
We thank Professor G. Pucacco for useful discussions on the use of the Weierstrass elliptic function in the SFM2 Hamiltonian.

\clearpage
\bigskip

\bibliography{sn-bibliography}


\begin{thebibliography}{32}
\ifx \bisbn   \undefined \def \bisbn  #1{ISBN #1}\fi
\ifx \binits  \undefined \def \binits#1{#1}\fi
\ifx \bauthor  \undefined \def \bauthor#1{#1}\fi
\ifx \batitle  \undefined \def \batitle#1{#1}\fi
\ifx \bjtitle  \undefined \def \bjtitle#1{#1}\fi
\ifx \bvolume  \undefined \def \bvolume#1{\textbf{#1}}\fi
\ifx \byear  \undefined \def \byear#1{#1}\fi
\ifx \bissue  \undefined \def \bissue#1{#1}\fi
\ifx \bfpage  \undefined \def \bfpage#1{#1}\fi
\ifx \blpage  \undefined \def \blpage #1{#1}\fi
\ifx \burl  \undefined \def \burl#1{\textsf{#1}}\fi
\ifx \doiurl  \undefined \def \doiurl#1{\url{https://doi.org/#1}}\fi
\ifx \betal  \undefined \def \betal{\textit{et al.}}\fi
\ifx \binstitute  \undefined \def \binstitute#1{#1}\fi
\ifx \binstitutionaled  \undefined \def \binstitutionaled#1{#1}\fi
\ifx \bctitle  \undefined \def \bctitle#1{#1}\fi
\ifx \beditor  \undefined \def \beditor#1{#1}\fi
\ifx \bpublisher  \undefined \def \bpublisher#1{#1}\fi
\ifx \bbtitle  \undefined \def \bbtitle#1{#1}\fi
\ifx \bedition  \undefined \def \bedition#1{#1}\fi
\ifx \bseriesno  \undefined \def \bseriesno#1{#1}\fi
\ifx \blocation  \undefined \def \blocation#1{#1}\fi
\ifx \bsertitle  \undefined \def \bsertitle#1{#1}\fi
\ifx \bsnm \undefined \def \bsnm#1{#1}\fi
\ifx \bsuffix \undefined \def \bsuffix#1{#1}\fi
\ifx \bparticle \undefined \def \bparticle#1{#1}\fi
\ifx \barticle \undefined \def \barticle#1{#1}\fi
\bibcommenthead
\ifx \bconfdate \undefined \def \bconfdate #1{#1}\fi
\ifx \botherref \undefined \def \botherref #1{#1}\fi
\ifx \url \undefined \def \url#1{\textsf{#1}}\fi
\ifx \bchapter \undefined \def \bchapter#1{#1}\fi
\ifx \bbook \undefined \def \bbook#1{#1}\fi
\ifx \bcomment \undefined \def \bcomment#1{#1}\fi
\ifx \oauthor \undefined \def \oauthor#1{#1}\fi
\ifx \citeauthoryear \undefined \def \citeauthoryear#1{#1}\fi
\ifx \endbibitem  \undefined \def \endbibitem {}\fi
\ifx \bconflocation  \undefined \def \bconflocation#1{#1}\fi
\ifx \arxivurl  \undefined \def \arxivurl#1{\textsf{#1}}\fi
\csname PreBibitemsHook\endcsname

\bibitem{henrard1983second}
\begin{barticle}
\bauthor{\bsnm{Henrard}, \binits{J.}},
\bauthor{\bsnm{Lemaitre}, \binits{A.}}:
\batitle{A second fundamental model for resonance}.
\bjtitle{Celestial mechanics}
\bvolume{30}(\bissue{2}),
\bfpage{197}--\blpage{218}
(\byear{1983})
\end{barticle}
\endbibitem

\bibitem{eftpaez2023}
\begin{bchapter}
\bauthor{\bsnm{Efthymiopoulos}, \binits{C.}},
\bauthor{\bsnm{Paez}, \binits{R.I.}}:
\bctitle{Arnold diffusion and nekhoroshev theory}.
In: \bbtitle{New Frontiers of Celestial Mechanics: Theory and Applications: I-CELMECH Training School, Milan, Italy, February 3--7, 2020},
pp. \bfpage{163}--\blpage{207}.
\bpublisher{Springer}, 
(\byear{2023})
\end{bchapter}
\endbibitem

\bibitem{chirikov1979universal}
\begin{barticle}
\bauthor{\bsnm{Chirikov}, \binits{B.V.}}:
\batitle{A universal instability of many-dimensional oscillator systems}.
\bjtitle{Physics reports}
\bvolume{52}(\bissue{5}),
\bfpage{263}--\blpage{379}
(\byear{1979})
\end{barticle}
\endbibitem

\bibitem{chierchia1994drift}
\begin{bchapter}
\bauthor{\bsnm{Chierchia}, \binits{L.}},
\bauthor{\bsnm{Gallavotti}, \binits{G.}}:
\bctitle{Drift and diffusion in phase space}.
In: \bbtitle{Annales de l'IHP Physique Th{\'e}orique},
vol. \bseriesno{60},
pp. \bfpage{1}--\blpage{144}
(\byear{1994})
\end{bchapter}
\endbibitem

\bibitem{delshams2006geometric}
\begin{bbook}
\bauthor{\bsnm{Delshams}, \binits{A.}},
\bauthor{\bparticle{De~la} \bsnm{Llave}, \binits{R.}},
\bauthor{\bsnm{Seara}, \binits{T.M.}}:
\bbtitle{A Geometric Mechanism for Diffusion in Hamiltonian Systems Overcoming the Large Gap Problem: Heuristics and Rigorous Verification on a Model: Heuristics and Rigorous Verification on a Model}
vol. \bseriesno{179}.
\bpublisher{American Mathematical Soc.}, \blocation{???}
(\byear{2006})
\end{bbook}
\endbibitem

\bibitem{delshams2008geometric}
\begin{barticle}
\bauthor{\bsnm{Delshams}, \binits{A.}},
\bauthor{\bsnm{De~La~Llave}, \binits{R.}},
\bauthor{\bsnm{Seara}, \binits{T.M.}}:
\batitle{Geometric properties of the scattering map of a normally hyperbolic invariant manifold}.
\bjtitle{Advances in Mathematics}
\bvolume{217}(\bissue{3}),
\bfpage{1096}--\blpage{1153}
(\byear{2008})
\end{barticle}
\endbibitem

\bibitem{guzzo2020semi}
\begin{barticle}
\bauthor{\bsnm{Guzzo}, \binits{M.}},
\bauthor{\bsnm{Efthymiopoulos}, \binits{C.}},
\bauthor{\bsnm{Paez}, \binits{R.I.}}:
\batitle{Semi-analytic computations of the speed of arnold diffusion along single resonances in a priori stable hamiltonian systems}.
\bjtitle{Journal of Nonlinear Science}
\bvolume{30}(\bissue{3}),
\bfpage{851}--\blpage{901}
(\byear{2020})
\end{barticle}
\endbibitem

\bibitem{arnold1964instability}
\begin{barticle}
\bauthor{\bsnm{Arnold}, \binits{V.I.}}:
\batitle{Instability of dynamical systems with several degrees of freedom}.
\bjtitle{Doklady Akademii Nauk SSSR}
\bvolume{156},
\bfpage{9}--\blpage{12}
(\byear{1964})
\end{barticle}
\endbibitem

\bibitem{LandauTeller1936}
\begin{barticle}
\bauthor{\bsnm{Landau}, \binits{L.}},
\bauthor{\bsnm{Teller}, \binits{E.}}:
\batitle{On the theory of sound dispersion}.
\bjtitle{Physik Zeitschrift der Sowjetunion}
\bvolume{10},
\bfpage{34}--\blpage{38}
(\byear{1936})
\end{barticle}
\endbibitem

\bibitem{jeans1903xxxv}
\begin{barticle}
\bauthor{\bsnm{Jeans}, \binits{J.}}:
\batitle{Xxxv. on the vibrations set up in molecules by collisions}.
\bjtitle{The London, Edinburgh, and Dublin Philosophical Magazine and Journal of Science}
\bvolume{6}(\bissue{32}),
\bfpage{279}--\blpage{286}
(\byear{1903})
\end{barticle}
\endbibitem

\bibitem{jeans1905xi}
\begin{barticle}
\bauthor{\bsnm{Jeans}, \binits{J.}}:
\batitle{Xi. on the partition of energy between matter and {\ae}ther}.
\bjtitle{The London, Edinburgh, and Dublin Philosophical Magazine and Journal of Science}
\bvolume{10}(\bissue{55}),
\bfpage{91}--\blpage{98}
(\byear{1905})
\end{barticle}
\endbibitem

\bibitem{rapp1960complete}
\begin{barticle}
\bauthor{\bsnm{Rapp}, \binits{D.}}:
\batitle{Complete classical theory of vibrational energy exchange}.
\bjtitle{The Journal of Chemical Physics}
\bvolume{32}(\bissue{3}),
\bfpage{735}--\blpage{737}
(\byear{1960})
\end{barticle}
\endbibitem

\bibitem{benettin1993landau}
\begin{barticle}
\bauthor{\bsnm{Benettin}, \binits{G.}},
\bauthor{\bsnm{Carati}, \binits{A.}},
\bauthor{\bsnm{Sempio}, \binits{P.}}:
\batitle{On the landau-teller approximation for energy exchanges with fast degrees of freedom}.
\bjtitle{Journal of statistical physics}
\bvolume{73}(\bissue{1}),
\bfpage{175}--\blpage{192}
(\byear{1993})
\end{barticle}
\endbibitem

\bibitem{benettin1997conservation}
\begin{barticle}
\bauthor{\bsnm{Benettin}, \binits{G.}},
\bauthor{\bsnm{Carati}, \binits{A.}},
\bauthor{\bsnm{Fasso}, \binits{F.}}:
\batitle{On the conservation of adiabatic invariants for a system of coupled rotators}.
\bjtitle{Physica D: Nonlinear Phenomena}
\bvolume{104}(\bissue{3-4}),
\bfpage{253}--\blpage{268}
(\byear{1997})
\end{barticle}
\endbibitem

\bibitem{rosengren2015chaos}
\begin{barticle}
\bauthor{\bsnm{Rosengren}, \binits{A.J.}},
\bauthor{\bsnm{Alessi}, \binits{E.M.}},
\bauthor{\bsnm{Rossi}, \binits{A.}},
\bauthor{\bsnm{Valsecchi}, \binits{G.B.}}:
\batitle{Chaos in navigation satellite orbits caused by the perturbed motion of the moon}.
\bjtitle{Monthly Notices of the Royal Astronomical Society}
\bvolume{449}(\bissue{4}),
\bfpage{3522}--\blpage{3526}
(\byear{2015})
\end{barticle}
\endbibitem

\bibitem{daquin2016dynamical}
\begin{barticle}
\bauthor{\bsnm{Daquin}, \binits{J.}},
\bauthor{\bsnm{Rosengren}, \binits{A.J.}},
\bauthor{\bsnm{Alessi}, \binits{E.M.}},
\bauthor{\bsnm{Deleflie}, \binits{F.}},
\bauthor{\bsnm{Valsecchi}, \binits{G.B.}},
\bauthor{\bsnm{Rossi}, \binits{A.}}:
\batitle{The dynamical structure of the meo region: long-term stability, chaos, and transport}.
\bjtitle{Celestial Mechanics and Dynamical Astronomy}
\bvolume{124}(\bissue{4}),
\bfpage{335}--\blpage{366}
(\byear{2016})
\end{barticle}
\endbibitem

\bibitem{celletti2016study}
\begin{barticle}
\bauthor{\bsnm{Celletti}, \binits{A.}},
\bauthor{\bsnm{Gale{\c{s}}}, \binits{C.B.}}:
\batitle{A study of the lunisolar secular resonance $2\dot{\omega} + \dot{\Omega}= 0$}.
\bjtitle{Frontiers in Astronomy and Space Sciences}
\bvolume{3},
\bfpage{11}
(\byear{2016})
\end{barticle}
\endbibitem

\bibitem{gkolias2016order}
\begin{barticle}
\bauthor{\bsnm{Gkolias}, \binits{I.}},
\bauthor{\bsnm{Daquin}, \binits{J.}},
\bauthor{\bsnm{Gachet}, \binits{F.}},
\bauthor{\bsnm{Rosengren}, \binits{A.J.}}:
\batitle{From order to chaos in earth satellite orbits}.
\bjtitle{The Astronomical Journal}
\bvolume{152}(\bissue{5}),
\bfpage{119}
(\byear{2016})
\end{barticle}
\endbibitem

\bibitem{rossi2018redshift}
\begin{barticle}
\bauthor{\bsnm{Rossi}, \binits{A.}},
\bauthor{\bsnm{Colombo}, \binits{C.}},
\bauthor{\bsnm{Tsiganis}, \binits{K.}},
\bauthor{\bsnm{Beck}, \binits{J.}},
\bauthor{\bsnm{Rodriguez}, \binits{J.B.}},
\bauthor{\bsnm{Walker}, \binits{S.}},
\bauthor{\bsnm{Letterio}, \binits{F.}},
\bauthor{\bsnm{Dalla~Vedova}, \binits{F.}},
\bauthor{\bsnm{Schaus}, \binits{V.}},
\bauthor{\bsnm{Popova}, \binits{R.}}, \betal:
\batitle{Redshift: a global approach to space debris mitigation}.
\bjtitle{Aerospace}
\bvolume{5}(\bissue{2}),
\bfpage{64}
(\byear{2018})
\end{barticle}
\endbibitem

\bibitem{celletti2017analytical}
\begin{barticle}
\bauthor{\bsnm{Celletti}, \binits{A.}},
\bauthor{\bsnm{Gale{\c{s}}}, \binits{C.}},
\bauthor{\bsnm{Pucacco}, \binits{G.}},
\bauthor{\bsnm{Rosengren}, \binits{A.J.}}:
\batitle{Analytical development of the lunisolar disturbing function and the critical inclination secular resonance}.
\bjtitle{Celestial Mechanics and Dynamical Astronomy}
\bvolume{127}(\bissue{3}),
\bfpage{259}--\blpage{283}
(\byear{2017})
\end{barticle}
\endbibitem

\bibitem{daquin2021deep}
\begin{barticle}
\bauthor{\bsnm{Daquin}, \binits{J.}},
\bauthor{\bsnm{Legnaro}, \binits{E.}},
\bauthor{\bsnm{Gkolias}, \binits{I.}},
\bauthor{\bsnm{Efthymiopoulos}, \binits{C.}}:
\batitle{A deep dive into the $2g+h$ resonance: separatrices, manifolds and phase space structure of navigation satellites}.
\bjtitle{Celestial Mechanics and Dynamical Astronomy}
\bvolume{134}(\bissue{1}),
\bfpage{1}--\blpage{31}
(\byear{2022})
\end{barticle}
\endbibitem

\bibitem{Legnaro_Efthymiopoulos_2022}
\begin{botherref}
\oauthor{\bsnm{Legnaro}, \binits{E.}},
\oauthor{\bsnm{Efthymiopoulos}, \binits{C.}}:
A detailed dynamical model for inclination-only dependent lunisolar resonances. effect on the “eccentricity growth” mechanism.
Advances in Space Research,
0273117722006871
(2022).
\doiurl{10.1016/j.asr.2022.07.057}
\end{botherref}
\endbibitem

\bibitem{efthymiopoulos1997stickiness}
\begin{barticle}
\bauthor{\bsnm{Efthymiopoulos}, \binits{C.}},
\bauthor{\bsnm{Contopoulos}, \binits{G.}},
\bauthor{\bsnm{Voglis}, \binits{N.}},
\bauthor{\bsnm{Dvorak}, \binits{R.}}:
\batitle{Stickiness and cantori}.
\bjtitle{Journal of Physics A: Mathematical and General}
\bvolume{30}(\bissue{23}),
\bfpage{8167}
(\byear{1997})
\end{barticle}
\endbibitem

\bibitem{contopoulos1999destruction}
\begin{barticle}
\bauthor{\bsnm{Contopoulos}, \binits{G.}},
\bauthor{\bsnm{Harsoula}, \binits{M.}},
\bauthor{\bsnm{Voglis}, \binits{N.}},
\bauthor{\bsnm{Dvorak}, \binits{R.}}:
\batitle{Destruction of islands of stability}.
\bjtitle{Journal of Physics A: Mathematical and General}
\bvolume{32}(\bissue{28}),
\bfpage{5213}
(\byear{1999})
\end{barticle}
\endbibitem

\bibitem{contopoulos2008stickiness}
\begin{barticle}
\bauthor{\bsnm{Contopoulos}, \binits{G.}},
\bauthor{\bsnm{Harsoula}, \binits{M.}}:
\batitle{Stickiness in chaos}.
\bjtitle{International Journal of Bifurcation and Chaos}
\bvolume{18}(\bissue{10}),
\bfpage{2929}--\blpage{2949}
(\byear{2008})
\end{barticle}
\endbibitem

\bibitem{contopoulos2010stickiness}
\begin{barticle}
\bauthor{\bsnm{Contopoulos}, \binits{G.}},
\bauthor{\bsnm{Harsoula}, \binits{M.}}:
\batitle{Stickiness effects in conservative systems}.
\bjtitle{International Journal of Bifurcation and Chaos}
\bvolume{20}(\bissue{07}),
\bfpage{2005}--\blpage{2043}
(\byear{2010})
\end{barticle}
\endbibitem

\bibitem{whittaker2020course}
\begin{bbook}
\bauthor{\bsnm{Whittaker}, \binits{E.T.}},
\bauthor{\bsnm{Watson}, \binits{G.N.}}:
\bbtitle{A Course of Modern Analysis},
\bedition{4}th edn.
\bsertitle{Cambridge Mathematical Library}.
\bpublisher{Cambridge University Press}, \blocation{???}
(\byear{1996}).
\doiurl{10.1017/CBO9780511608759}
\end{bbook}
\endbibitem

\bibitem{cincotta2002arnold}
\begin{barticle}
\bauthor{\bsnm{Cincotta}, \binits{P.M.}}:
\batitle{Arnold diffusion: an overview through dynamical astronomy}.
\bjtitle{New Astronomy Reviews}
\bvolume{46}(\bissue{1}),
\bfpage{13}--\blpage{39}
(\byear{2002})
\end{barticle}
\endbibitem

\bibitem{cincotta2014chirikov}
\begin{barticle}
\bauthor{\bsnm{Cincotta}, \binits{P.M.}},
\bauthor{\bsnm{Efthymiopoulos}, \binits{C.}},
\bauthor{\bsnm{Giordano}, \binits{C.M.}},
\bauthor{\bsnm{Mestre}, \binits{M.F.}}:
\batitle{Chirikov and nekhoroshev diffusion estimates: bridging the two sides of the river}.
\bjtitle{Physica D: Nonlinear Phenomena}
\bvolume{266},
\bfpage{49}--\blpage{64}
(\byear{2014})
\end{barticle}
\endbibitem

\bibitem{cincotta2023estimation}
\begin{barticle}
\bauthor{\bsnm{Cincotta}, \binits{P.M.}},
\bauthor{\bsnm{Giordano}, \binits{C.M.}}:
\batitle{Estimation of diffusion time with the shannon entropy approach}.
\bjtitle{Physical Review E}
\bvolume{107}(\bissue{6}),
\bfpage{064101}
(\byear{2023})
\end{barticle}
\endbibitem

\bibitem{press2007numerical}
\begin{bbook}
\bauthor{\bsnm{Press}, \binits{W.H.}},
\bauthor{\bsnm{Teukolsky}, \binits{S.A.}},
\bauthor{\bsnm{Vetterling}, \binits{W.T.}},
\bauthor{\bsnm{Flannery}, \binits{B.P.}}:
\bbtitle{Numerical Recipes 3rd Edition: The Art of Scientific Computing}.
\bpublisher{Cambridge university press},
(\byear{2007})
\end{bbook}
\endbibitem

\bibitem{Gondelach_Armellin_Wittig_2019}
\begin{barticle}
\bauthor{\bsnm{Gondelach}, \binits{D.J.}},
\bauthor{\bsnm{Armellin}, \binits{R.}},
\bauthor{\bsnm{Wittig}, \binits{A.}}:
\batitle{On the predictability and robustness of galileo disposal orbits}.
\bjtitle{Celestial Mechanics and Dynamical Astronomy}
\bvolume{131}(\bissue{12}),
\bfpage{60}
(\byear{2019}).
\doiurl{10.1007/s10569-019-9938-9}
\end{barticle}
\endbibitem

\end{thebibliography}


\end{document}